\documentstyle[12pt]{article}

\begin{document}
\title{Formal solutions of stargenvalue equations}
\author{
Nuno Costa Dias\footnote{{\it ncdias@mail.telepac.pt}} \\
Jo\~ao Nuno Prata\footnote{{\it joao.prata@ulusofona.pt}} \\
{{\it Departamento de Matem\'{a}tica}} \\
{{\it Universidade Lus\'ofona de Humanidades e Tecnologias}} \\ {{\it Av. Campo Grande, 376,
1749-024 Lisboa, Portugal}}}
\date{}
\maketitle

\begin{abstract}
The formal solution of a general stargenvalue equation is presented, its properties studied and a geometrical interpretation given in terms of star-hypersurfaces in quantum phase space. Our approach deals with discrete and continuous spectra in a unified fashion and includes a systematic treatment of non-diagonal stargenfunctions. The formalism is used to obtain a complete formal solution of Wigner quantum mechanics in the Heisenberg picture and to write a general formula for the stargenfunctions of Hamiltonians quadratic in the phase space variables in arbitrary dimension. A variety of systems is then used to illustrate the former results.
\end{abstract}

\section{Introduction}

The Wigner formulation of quantum mechanics \cite{Weyl} - \cite{nuno6} has become an important field of research. This is probably due to the fact that
Wigner theory formulates quantum mechanics in terms of "classical-like" objects. Because of this, it is perceived by many as more intuitive than the standard operator formulation and has been used to successfully address a considerable number of problems in a variety of fields of research ranging from the semiclassical limit of quantum mechanics \cite{Lee1,Flato1,Lee2,Smith,nuno3}, quantum chaos \cite{latka,shin} and hybrid dynamics \cite{nuno2} to $M$-theory \cite{Fairlie2,Fairlie3,Witten,Pinzul}.

Possibly, the main advantage of Wigner quantum mechanics is conceptual and stems from its remarkable relation with classical statistical mechanics. The theory is formulated in terms of the phase space Wigner function which for one-dimensional dynamical systems reads:
\begin{equation}
F^W(q,p,t)=\frac{1}{2 \pi \hbar} W(|\psi(t)><\psi(t)|)= \frac{1}{2 \pi} \int dy e^{-i p y} < q+\frac{\hbar}{2} y|\psi(t)><\psi(t)| q-\frac{\hbar}{2} y>,
\end{equation}
and is the Weyl transform of the density matrix $|\psi(t)><\psi(t)|$. The Weyl map $W:\hat{\cal A} ({\cal H}) \to {\cal A}(T^*M)$ can also be used to obtain the phase space representation $A(q,p)=W(\hat A)$ of a generic operator $\hat A$ belonging to the algebra of linear operators $\hat{\cal A}({\cal H}) $ acting on the Hilbert space ${\cal H}$. Furthermore, using the Schr\"odinger equation and the Weyl map we can easily derive the dynamics of the Wigner function, \cite{Moyal}:
$$
\dot{F}^W(q,p;t)= [H( q, p),F^W ( q, p;t)]_M,
$$
where $H(q,p)=W(\hat H)$ is the Weyl symbol of the quantum Hamiltonian $\hat H$. The Moyal bracket,
$$
[A(q,p) ,B(q,p) ]_M= \frac{1}{i \hbar} \left( A(q,p)*B(q,p)-B(q,p)*A(q,p) \right),
$$
and the star product $*$ are both $\hbar$-deformations of the algebraic structures of classical mechanics (the standard product and the Poisson bracket, respectively) \cite{nuno6,Flato1}. They can be defined through the relations:
$$
W(\hat A \cdot \hat B)=W(\hat A) * W(\hat B), \quad
W\left (\frac{1}{i \hbar} [\hat A, \hat B] \right )=[W(\hat A) , W(\hat B)]_M, \quad \forall \hat A,\hat B \in \hat{\cal A},
$$
from which their explicit functional form follows immediately:
$$
A*B  =
A \exp \left\{\frac{i \hbar}{2} \left(\frac{ {\buildrel { \leftarrow}\over\partial}}{\partial q} \frac{ {\buildrel {
\rightarrow}\over\partial}}{\partial p} -  \frac{{\buildrel { \leftarrow}\over\partial}}{\partial p}  \frac{{\buildrel {
\rightarrow}\over\partial}}{\partial q} \right) \right\} B,\quad
[A,B]_M  =  \frac{2}{ \hbar} A \sin \left\{ \frac{\hbar}{2} \left(\frac{ {\buildrel { \leftarrow}\over\partial}}{\partial q} \frac{ {\buildrel {
\rightarrow}\over\partial}}{\partial p} -  \frac{{\buildrel { \leftarrow}\over\partial}}{\partial p}  \frac{{\buildrel {
\rightarrow}\over\partial}}{\partial q} \right) \right\} B,
$$
the derivatives ${\buildrel { \leftarrow}\over\partial}$ and ${\buildrel { \rightarrow}\over\partial}$ acting on $A$ and $B$, respectively. With these structures the
Weyl map becomes an isomorphism between the Lie algebra of quantum operators $(\hat{\cal A}({\cal H}),\cdot,[\quad ,\quad])$ and the Lie algebra of phase space functionals $({\cal A}(T^*M),*,[\quad ,\quad]_M)$.

This structure yields basic physical predictions through the average value,
$$
<A(q,p;t)>= \int dq dp A(q,p) F^W(q,p;t),
$$
and the marginal probability functionals,
\begin{equation}
P(q(t)=q_0)= \int dp  F^W (q_0,p;t) \quad \mbox{and} \quad P(p(t)=p_0)= \int dq   F^W(q,p_0;t),
\end{equation}
both displaying an impressive similarity with their analogues in classical mechanics.

To produce more general predictions one has to introduce the stargenvalue equation, which is the Weyl transform of the eigenvalue equation \cite{Fairlie1,nuno5}. For a one-dimensional system this equation reads:
\begin{equation}
W(\hat A |a><a|) = W( a |a><a|) \Longleftrightarrow
A( q, p) * g_a( q, p) = a g_a( q, p),
\end{equation}
where $|a>$ is a general eigenstate of $\hat A$ with associated (non-degenerate) eigenvalue $a$ and $g_a(q,p)= W(|a><a|)$ is the {\it stargenfunction} associated to the same eigenvalue. The stargenfunctions $g_a(q,p)$ can then be used to obtain the probabilities that a measurement of a generic observable $A(q,p)$ yield the value $a$:
$$
P(A( q, p;t)=a)= \int dq dp \hspace{0.2 cm} g_a( q, p) F^W( q,p;t),
$$
thus generalizing eq.(2).

The solutions of the stargenvalue equation can literally be connected to all relevant structures of the Wigner theory and it is thus not surprising that this equation has been extensively studied in the past. Its properties were systematically described in \cite{Flato1,Fairlie2,Fairlie1,nuno5,Kundt,Dahl1} and solutions (both analytical and numerical) for several particular systems were presented in \cite{Fairlie2,Fairlie1,Dahl1,Hug}. There is also a vast literature on the subject of determining the semiclassical behavior of these stargenfunctions. In his seminal work \cite{Berry}, Berry derived the general semiclassical form for the eigenstates of a generic observable, using the Wigner function associated with the corresponding WKB-wavefunctions. Ozorio de Almeida and Hannay \cite{Almeida1} generalized this approach to higher-dimensional systems. More recently, the dynamics of these semiclassical states has been studied in depth, \cite{Almeida2,Osborn}. In particular, the authors addressed the issue of whether it makes sense to replace the Moyal bracket by the Poisson bracket for these semiclassical states. More details about these issues can be found in \cite{Almeida3}.

In this paper we revisit the problem of obtaining exact solutions for a generic stargenvalue equation.
A general formal solution will be presented both for continuous and discrete spectra. Our approach includes a systematic treatment of non-diagonal stargenfunctions and higher-dimensional systems and leads to 1) a geometrical interpretation of the stargenfunctions in terms of what will be named {\it star hypersurfaces} in quantum phase space, 2) a complete formal solution of Wigner quantum mechanics in the Heisenberg picture and 3) a general expression for the stargenfunctions of Hamiltonians quadratic in the phase space variables in arbitrary dimension. In this context
we also dwell on the possibility of obtaining integral representations for polynomials associated with certain solutions of the time-independent Schr\"odinger equation.

This last result integrates and generalizes several results previously presented in the literature, regarding solutions of the stargenvalue equation for specific quadratic Hamiltonians \cite{Flato1}. It also proves the efficiency of the formal solution as a calculation tool in a variety of systems.

However, it is important to emphasize that the main advantage of our approach is conceptual. The formal solution and the Heisenberg picture provide universal rules and formulae that are applicable to very disparate situations.
They played a key role in relating the De Broglie-Bohm and the Wigner formulations of quantum mechanics (cf.\cite{nuno1,nuno9}) and were systematically used to study the semiclassical limit of quantum mechanics in \cite{Berry,Almeida1}. We also believe they might be particularly relevant for the field of constrained dynamical systems \cite{nuno7} (and in particular for general parameterized systems) where the characterization of the physical space of quantum states is the main conceptual and technical issue of any quantization program \cite{Henneaux}.

This paper is organized as follows: in section 2 we discuss the general nature of projectors in standard operator quantum mechanics and use the Weyl map to derive the formal solution of a general stargenvalue equation. We also prove that stargenfunctions are an $\hbar$-deformation of the Dirac distribution and introduce the concept of star hypersurface.
In section 3 we formulate Wigner quantum mechanics in the Heisenberg picture and present its complete formal solution. In section 4 we derive the explicit form of the stargenfunctions of a generic quadratic Hamiltonian in arbitrary dimension. Sections 5,6 and 7 illustrate the previous results through three simple examples: one-dimensional Hamiltonians with discrete (section 5 - simple harmonic oscillator) and continuous (section 6 - linear potential) spectra and a two-dimensional example (section 7 - two-dimensional harmonic oscillator), where we compute the simultaneous stargenfunctions (both diagonal and non-diagonal) of the energy and the $z$-component of angular momentum. Finally, in section 8, we present our conclusions.

\section{Projectors and stargenfunctions}

The cases of continuous and discrete spectra will be considered separately.

\subsection{Continuous spectrum}

Let us start with a one-dimensional system and consider a hermitian operator $\hat A$ with non-degenerate continuous spectrum. Let $|a>$ be the general eigenstate of $\hat A$ with associated eigenvalue $a$.
The explicit form of the projector $|a><a|$, which will be designated by $\hat {\Delta} (\hat A -a)$, is given by:
\begin{equation}
|a><a|= \hat {\Delta} (\hat A -a)=  \frac{1}{2 \pi}\int dk e^{ik(\hat A-a)}.
\end{equation}
To prove this result let us introduce two general states:
$|\phi>$ and $|\psi>$. Using the representation of $\hat A$ we have:
\begin{eqnarray}
<\psi|\hat {\Delta} (\hat A -a)|\phi> & = & \frac{1}{2 \pi}\int da' da'' <\psi|a'><a'|\int dk e^{ik (\hat A-a)}|a''><a''|\phi> \nonumber \\
& = & \frac{1}{2 \pi}\int da' da'' <\psi|a'> \delta (a''-a') \int dk e^{ik (a''-a)}<a''|\phi> \nonumber \\
& = & \int da' <\psi|a'> \delta( a'-a) <a'|\phi> = <\psi|a><a|\phi>,
\end{eqnarray}
from which the identity (4) follows immediately.

Let us now consider a dynamical system of arbitrary (but finite) dimension. Let $\hat A$ be an observable with eigenstates $|a, \vec z>$ such that $\hat A |a, \vec z> = a |a, \vec z>$ and $\vec z$ is a array of continuous degeneracy indices. It is then also true that:
\begin{equation}
\int d \vec z \,|a, \vec z><a, \vec z\,| =  \frac{1}{2 \pi}\int dk e^{ik(\hat A-a)} = \hat {\Delta} (\hat A -a).
\end{equation}
In fact:
\begin{eqnarray}
&& \frac{1}{2 \pi} <\psi|\int dk e^{ik(\hat A - a)}|\phi> = \nonumber \\
& = & \frac{1}{2 \pi} \int \int da' d \vec z\,' \int \int da'' d \vec z\,''
<\psi|a' ,\vec z\,'><a' ,\vec z\,'|\int dk e^{ik(\hat A -a)}|a'' ,\vec z\,''><a'' ,\vec z\,''| \phi> = \nonumber \\
& = & \int \int da' d \vec z\,' \int \int da'' d \vec z\,'' \delta (a''-a)
<\psi|a', \vec z\,'><a' ,\vec z\,'|a'' ,\vec z\,''><a'', \vec z\,''| \phi>= \nonumber \\
& = & \int d \vec z\,' <\psi|a ,\vec z\,'><a ,\vec z\,'| \phi>,
\end{eqnarray}
which proves that the functional form of the projector $\hat {\Delta} (\hat A -a)$ is always given by eq.(4) independently of the dimension of the system.

We now consider a two dimensional system. Let $\hat B$ be a second observable also with continuous spectrum and such that $[\hat A, \hat B] =0$. The set of simultaneous eigenvectors $|a,b>$ (such that $\hat A |a,b>= a |a,b>$ and $\hat B |a,b>= b|a,b>$) spans the Hilbert space of the system. The projector $|a,b><a,b|$ is given by:
$
|a,b><a,b|= \hat {\Delta } (\hat A-a) \hat {\Delta }(\hat B -b)$, a result that follows from:
\begin{eqnarray}
& & \hat {\Delta } (\hat A-a) \hat {\Delta }(\hat B -b)  =  \int db' \int da' |a, b'><a, b'|a', b><a',b| =  \nonumber \\
& & = \int db' \int da' |a, b'><a' , b| \delta (a'-a) \delta (b-b')  =  |a,b><a,b|.
\end{eqnarray}
The generalization to higher dimensions is straightforward: let $\{\hat A_i, i=1..n \}$ be a complete set of commuting observables displaying continuous spectra. The set of eigenvectors $|a_1,..,a_i,..,a_n>$ (such that $\hat A_i|a_1,..,a_i,..,a_n> = a_i
|a_1,..,a_i,..,a_n>$, $\forall i=1..n$ ) spans the Hilbert space of the system. The projector
$|a_1,..,a_i,..,a_n><a_1,..,a_i,..,a_n|$ reads:
\begin{equation}
|a_1,..,a_i,..,a_n><a_1,..,a_i,..,a_n| = \hat {\Delta } (\hat A_1-a_1) ..\hat {\Delta } (\hat A_i-a_i) ...\hat {\Delta } (\hat A_n-a_n) .
\end{equation}
Notice that it is crucial to our approach that the set of operators $\{\hat A_i \} $ forms a complete set of commuting observables. If this is not the case (like for the Hamiltonian of a classical chaotic system) then the former resolution of the problem is not valid. From now on we will always assume that $\{\hat A_i \}$ is a complete set of commuting observables.

To proceed we address the problem of obtaining the explicit functional form of the non-diagonal elements $|\vec b><\vec a|$. We start by considering the one-dimensional case and introduce the "translation" operator: $\hat T (\lambda)|a>=|a+\lambda>$. If $\hat B$ is such that $[\hat A,\hat B]=i\hbar$ then $\hat T (\lambda)=\exp (- i/\hbar \lambda \hat B)$, and thus:
\begin{equation}
|b><a|= \hat T (b-a) |a><a| = \frac{1}{2 \pi}\int dk e^{- \frac{i}{\hbar} (b-a) \hat B}  e^{ik (\hat A -a)} = \frac{1}{2 \pi}\int dk e^{- \frac{i}{\hbar} (b-a) \hat B + ik (\hat A -\frac{a+b}{2} )},
\end{equation}
where in the last step, we used the Baker-Campbell-Hausdorff formula.
The operator $|b><a|$ will be denoted by $\hat{\Delta} (\hat A,b,a)$. It is trivial to check that $\hat{\Delta} (\hat A,b,a)$ satisfies
$ \hat{\Delta} (\hat A,b,a) \hat{\Delta} (\hat A,c,d) = \delta(a-c) \hat{\Delta} (\hat A,b,d)$ and so
$\int da \hat{\Delta} (\hat A,b,a)\hat{\Delta} (\hat A,c,d)= \hat{\Delta} (\hat A,b,d)$.
A small remark is in order: the operator $\hat B$ is not completely specified by the commutation relation $[\hat A,\hat B]=i\hbar $. For instance $\hat B$ and $\hat B+F(\hat A)$ satisfy the same commutation relations for generic functional $F$. Different operators $\hat B$ lead to different specifications of the non-diagonal element $|b><a|$ (cf.(10)). If $\hat B$ is required to be hermitian all possible elements $|b><a|$ (for different $\hat B$ and fixed $b,a$) are related by a phase factor which is an exclusive function of $\lambda =b-a$. This freedom has no effect on the physical predictions of the theory. Moreover our future results are equally valid for all possible choices of $\hat B$ in
eq.(10) (providing they satisfy the proper commutation relations) and thus from now on, we shall assume that $\hat B$ is fully specified.

These results can be generalized to higher dimensions. We introduce the translation operators $\hat T_i$ such that:
$\hat T_i (\lambda_i) |a_1,..,a_i,..,a_n>=|a_1,..,a_i+\lambda_i,..,a_n> , \quad \forall i=1..n $.
If $\{\hat B_i, i=1..n\}$ is another complete set of mutually commuting observables satisfying $[\hat A_i, \hat B_j] = i\hbar \delta_{ij}$ then $\hat T_i (\lambda_i) = \exp (- i/\hbar \lambda_i \hat B_i)$,  and the general non-diagonal projector reads:
\begin{eqnarray}
&& |a_1',..,a_n'><a_1,..,a_n|  =  \hat T_1(a_1'-a_1) ... \hat T_n(a_n'-a_n) |a_1,..,a_n><a_1,..,a_n| = \nonumber \\
& & =  \exp \left\{- i/\hbar \{(a_1'-a_1) \hat B_1 + ...+(a_n'-a_n) \hat B_n \}\right\}
\hat{\Delta} (\hat A_1 -a_1)...\hat{\Delta} (\hat A_n-a_n) = \nonumber \\
& & =  \hat{\Delta} (\hat A_1,a_1',a_1)...\hat{\Delta} (\hat A_n,a_n',a_n).
\end{eqnarray}
Moreover, it is easy to prove that the {\it single} projector,
\begin{equation}
|a_i'><a_i|= \int da_1...\int da_{i-1} \int da_{i+1}...\int da_n |a_1,..,a_i',..,a_n><a_1,..,a_i,..,a_n|
\end{equation}
is given by:
\begin{equation}
|a_i'><a_i|= \hat T(a_i'-a_i) |a_i><a_i| = \hat{\Delta} ( \hat A_i, a_i',a_i)
\end{equation}
where $|a_i><a_i|$ is given by eq.(6). Therefore the non-diagonal single projector (12) is of the general form (10) in any dimension.

Finally, let us consider the problem of determining the general solution of the stargenvalue equation. We first consider the one-dimensional case. The most general stargenvalue equation for an arbitrary Weyl symbol $A$ is written \cite{Fairlie1}:
\begin{equation}
A * g_{ba} =b g_{ba} \quad \mbox{and} \quad g_{ba}*A = ag_{ba}
\end{equation}
and is the Weyl transform of the corresponding eigenvalue equation in the density matrix formulation of quantum mechanics:
$\hat A |\psi><\phi| = b |\psi><\phi|$ and $ |\psi><\phi| \hat A = a |\psi><\phi|$.
The general solution of this last equation is given by the projector (10) and thus the Weyl transform of (10) is the general solution of eq.(14):
\begin{equation}
\Delta_*(A(q,p),b,a)=\frac{1}{2 \pi}\int dk e_*^{- \frac{i}{\hbar} (b-a) B(q,p)} * e_*^{ik (A(q,p) -a)}= \frac{1}{2 \pi}\int dk e_*^{- \frac{i}{\hbar} (b-a) B(q,p) + ik (A(q,p) - \frac{a+b}{2}  )}  ,
\end{equation}
the star exponential $e_*$ being defined by:
\begin{equation}
e_*^{A(q,p)}=\sum_{n=0}^{\infty} \frac{1}{n!} A(q,p)^{*n},
\end{equation}
where $A(q,p)^{*n}$ is the $n$-fold starproduct of $A(q,p)$ and $A= A(q,p)=W(\hat A)$, $B= B(q,p)=W(\hat B)$. Furthermore, if $a=b$ then eq.(15) yields the diagonal element which is of the form:
\begin{equation}
\Delta_*(A(q,p),a,a)= \Delta_*(A(q,p)-a)= \frac{1}{2 \pi}\int dk  e_*^{ik (A(q,p) -a)}.
\end{equation}
In section 2.3 we will prove that this object is formally a $\hbar$-deformation of the Dirac delta function: $\Delta_*(A(q,p)-a)=\delta(A(q,p)-a)+{\cal O}(\hbar )$, the full identity being valid for those observables satisfying $A^{*n}=A^n$.

The generalization to $n$-dimensional systems is easily carried out if one uses the formal expression of the $n$-dimensional projector (given by eq.(11)) as a starting point:
\begin{eqnarray}
W(|a_1',..,a_n'><a_1,..,a_n|) & = & \Delta_* ( A_1,a_1',a_1)*..*\Delta_* (A_n,a_n',a_n) \nonumber \\
& = & \Delta_* ( A_1,..,A_n;a_1',..,a_n';a_1,..,a_n),
\end{eqnarray}
where $\Delta_* ( A_i,a_i',a_i)= W\{\hat{\Delta} (\hat A_i,a_i',a_i)\}$ is the single stargenfunction also given by (15) this time with $A=A(q_1,..,q_n,p_1,..,p_n)$ and $B=B(q_1,..,q_n,p_1,..,p_n)$. Moreover the notation $\Delta_* ( A_1,..,A_n;a_1',..,a_n';a_1,..,a_n)$ was introduced to designate the most general $n$-dimensional stargenfunction.
From the previous discussion one is led to the conclusion that:
\begin{eqnarray}
A_i* \Delta_* ( A_1,...,A_n;a_1',...,a_n';a_1,...,a_n) & = & a_i' \Delta_* ( A_1,...,A_n;a_1',...,a_n';a_1,...,a_n), \nonumber \\
\Delta_* ( A_1,...,A_n;a_1',...,a_n';a_1,...,a_n)* A_i & = & a_i\Delta_* ( A_1,...,A_n;a_1',...,a_n';a_1,...,a_n),
\end{eqnarray}
an identity that is valid for all $i=1..n$ and that can be checked explicitly by substitution of eqs.(15,18) in eq.(19). In particular, if $a_i=a_i'$ then $\Delta_* ( A_1,..,A_i,..,A_n;a_1',..,a_i,..,a_n';a_1,..,a_i,..,a_n)$ is one of the $a_i$-left and -right stargenfunctions of the observable $A_i$.
Furthermore, notice that the single stargenfunction $\Delta_* ( A_i,a_i',a_i)$ also satisfies the former stargenvalue equation (this time just for a single value of $i$). In fact, the relation between the single and the $n$-dimensional stargenfunctions is very appealing: on the one hand they are related by eq.(18) and on the other hand, from eq.(12) they also satisfy:
\begin{equation}
\Delta_* ( A_i,a_i',a_i) = \int da_1 ...\int da_{i-1} \int da_{i+1} ..\int da_n
\Delta_* ( A_1,..,A_i,..,A_n;a_1,..,a_i',..,a_n;a_1,..,a_i,..,a_n)
\end{equation}
We conclude that in the context of Wigner quantum mechanics the $n$-dimensional stargenfunctions can always be constructed from the single ones and therefore we shall henceforth focus on the one-dimensional case only.

\subsection{Discrete spectrum}

The case of discrete spectrum is slightly more involved.
The first step will be to introduce a "continuous like notation" allowing for a formulation of the discrete spectrum case in terms of the continuous spectrum formalism. Using this notation the entire set of results of the last section can be easily translated to the discrete spectrum case.

Let $\hat A$ be an observable with discrete spectrum and let $\{|a_n>\}$ form a complete orthonormal set of eigenstates of $\hat A$ with associated non-degenerate eigenvalues $a_n$. We now introduce the "continuous like notation" by defining the continuous projector:
\begin{equation}
|a><a|=\sum_n \delta(a-a_n) |a_n><a_n|,
\end{equation}
which is identically zero for all values of $a$ that do not belong to the spectrum of $\hat A$.
The intention is to use the matrix elements $|a><a|$ and the continuous spectrum formalism to reproduce the discrete spectrum results. We start by proving that $\{|a><a|,\quad a \in {\cal R} \}$ is a complete set of projectors. Let then $|\phi>$ and $|\psi>$ be two general states:
\begin{eqnarray}
<\phi|\int da |a><a| \psi> & = & \int da \sum_n \delta (a-a_n) <\phi|a_n><a_n|\psi>\nonumber \\
&=& \sum_n <\phi|a_n><a_n|\psi> =<\phi|\psi>,
\end{eqnarray}
and thus $\int da |a><a|=1$. Moreover:
\begin{eqnarray}
|a'><a'|a><a|& = &  \sum_{n,m} \delta (a'-a_n)\delta (a-a_m) |a_n><a_n|a_m><a_m| = \nonumber \\
& = & \sum_{n,m} \delta (a'-a_n)\delta (a-a_m) \delta_{n,m} |a_n><a_m| \nonumber \\
& = & \delta (a-a') \sum_n \delta (a-a_n) |a_n><a_n| =
\delta (a-a') |a><a|,
\end{eqnarray}
and thus, as expected $ |a><a|$ is a well defined projector. Finally, we consider the probability distribution resulting from using the continuous spectrum formalism. If $|\psi>$ is the state of the system then $ {\cal P}(A=a)= \mbox{tr}(|\psi><\psi||a><a|)=\sum_n \delta(a-a_n) |<\psi|a_n>|^2$
and thus:
\begin{eqnarray}
P(A=a) & = & \lim_{\epsilon \to 0} \int_{a-\epsilon}^{a+\epsilon} da' {\cal P} (A=a') \\
& = & \lim_{\epsilon \to 0} \int_{a-\epsilon}^{a+\epsilon} da'\sum_n \delta(a'-a_n) |<\psi|a_n>|^2=
\left\{ \begin{array}{l}
0 \quad \mbox{if} \quad a \not=a_n, \; \forall n \\
\\
|<\psi|a_n>|^2 \quad \mbox{if} \quad \exists n: a=a_n
\end{array}
\right. \nonumber
\end{eqnarray}
as it should.

The primary result concerning the stargenfunctions of $\hat A$ is that the projector $|a><a|$ (21) is also given by eq.(4), that is:
\begin{equation}
|a><a|=\sum_n \delta (a-a_n)|a_n><a_n| = \hat{\Delta}(\hat A - a).
\end{equation}
To see this explicitly we introduce two general states $|\psi>$ and $|\phi>$ and proceed as in (5):
\begin{eqnarray}
<\phi|\hat{\Delta} (\hat A-a) |\psi> &=&
\frac{1}{2 \pi } \sum_{n , m} <\phi |a_n><a_n| \int dk \exp \{ {ik (\hat A -a) } \} |a_m><a_m|\psi> =\nonumber \\
& = & \sum_{n , m} \delta (a_m-a) <\phi |a_n><a_n| a_m><a_m|\psi> = \nonumber \\
& = & \sum_{m} \delta (a_m-a) <\phi | a_m><a_m|\psi> = <\phi|a><a|\psi>.
\end{eqnarray}
The straightforward corollary being that $\hat{\Delta}(\hat A-a)=0$ if $a\not=a_n$ for all $n$. The generalization to higher dimensions follows exactly the same steps as in the continuous case and the discrete spectrum stargenfunctions also satisfy eqs.(6,8,9).

The non-diagonal elements can also be easily obtained if one knows the explicit form of the translation operator $\hat T(\lambda)$. Notice that in the discrete spectrum case this operator is not of the form used in eq.(10), given the fact that there is no operator $\hat B$ satisfying $[\hat A,\hat B]=i\hbar$. For instance, for the harmonic oscillator and for $\hat A= \hat H$ we have $\hat T(\lambda =nw\hbar)=\hat{\alpha}^n$ and $\hat T(\lambda \not=nw\hbar)=0$, where $\hat{\alpha}$ is the creation or the destruction operator and $n\in {\cal Z}$.

In general, let $a_m < a_n$ be two eigenvalues of $\hat A$, and define $\lambda_{n,m} = a_n - a_m$. The translation operator $\hat T (\lambda_{n,m})$ is such that:
\begin{equation}
\left[\hat A, \hat T (\lambda_{n,m}) \right]= \lambda_{n,m}\hat T (\lambda_{n,m}).
\end{equation}
We then have:
$$
\begin{array}{c}
\hat A \left( \hat T (\lambda_{n,m}) |a_m> \right) = \hat T (\lambda_{n,m}) \hat A |a_m> + \left[\hat A, \hat T (\lambda_{n,m}) \right]|a_m>=\\
= \left(a_m + \lambda_{n,m} \right)  \hat T (\lambda_{n,m}) |a_m> = a_n \hat T (\lambda_{n,m}) |a_m >,
\end{array}
$$
which means that $\hat T (\lambda_{n,m}) |a_m>$ is an eigenstate of $\hat A$ with eigenvalue $a_n$. There is a wide range of operators that fall into this classification. In fact, let us consider the classical semi-simple Lie Algebras ($A_n$, $B_n$, $C_n$, $D_n$, $E_6,\cdots$) and construct the Chevalley canonical form \cite{Cornwell} of the generators: $\left\{E_{\alpha}, E_{- \alpha}, H_{\alpha} \right\}$. Here $\alpha \in \Delta$ represent the set of simple roots, and $H_{\alpha}$ are the corresponding elements in the Cartan subalgebra. Then, for any given $l$-dimensional representation of the algebra, the operators $E_{\pm \alpha}$ work as ladder operators in the sense of equation (27). The most celebrated example is the angular momentum associated with the $su(2)$ Lie algebra, where $E_{\pm \alpha}$ correspond to the ladder operators $J_{\pm} = J_x \pm i J_y$, which implement translations in the spectrum of $H_{\alpha}$, or $J_z$.

Once we have found an operator $\hat T (\lambda_{n,m} )$ satisfying eq.(27), we get:
$|a_n><a_m| \propto \hat T(\lambda_{n,m})|a_m><a_m|$. In the "continuous spectrum notation" the general non-diagonal element is then $\hat T (\lambda_{n,m}) \hat{\Delta} (\hat A-a_m)$
and yields (using the Weyl map) the general stargenfunction of $A(q,p)$, (let $T(\lambda_{n,m})=W(\hat T(\lambda_{n,m}))$:
\begin{equation}
\Delta_*(A,a_n,a_m)\equiv T(\lambda_{n,m}) * \frac{1}{2 \pi}\int dk e_*^{ik(A(q,p)-a_m)}.
\end{equation}
This is the most general formula for non-diagonal elements and it is applicable both to continuous as well as discrete spectra.

A slightly more useful formula can nevertheless be derived, provided the steps $\lambda_{n,m}$ are constant, i.e. $|a_{n+1}-a_n |= \lambda$, $\forall n$. In that case, let $\hat T=\hat T (\lambda)$ be the translation operator. The spectrum is defined by $a_m = a_n + (m-n) \lambda, \hspace{0.3 cm} (m, n \in I)$, where $I$ is some set of integers. We can write:
\begin{equation}
|a_m> = |a_n + (m-n) \lambda> = \beta_{n,m} \hat T^{m-n} |a_n>,
\end{equation}
where $\beta_{n,m}$ is some normalization constant. Consequently,
\begin{equation}
|a+ n \lambda> < a| = \hat T^n |a><a| = \sum_{k \in I} \delta (a -a_k)  \frac{1}{\beta_{k,n+k}} | a_k + n \lambda>< a_k|.
\end{equation}
The Wigner functions associated with the non-diagonal elements of the density matrix, can thus be read off from the previous formula. In this respect it will prove useful to compute the following product of exponentials: $e^{\alpha \hat T} \cdot e^{ik \hat A}$. Since the commutator $\left[\hat T, \hat A \right]$ is proportional to $\hat T$, we conclude that all commutators appearing in the Baker-Campbell-Hausdorff formula either vanish or are proportional to $\hat T$. Let us then write:
\begin{equation}
e^{\alpha \hat T} \cdot e^{i k \hat A} = e^{i k \hat A + \alpha \mu (k) \hat T},
\end{equation}
where $\mu (k)$ is some function of $k$ yet to be determined. If we expand the exponentials in the previous equation and equate powers of $\hat T$ and $\hat A$, we conclude that:
\begin{equation}
\mu (k) = \frac{i k \lambda}{e^{ik \lambda} -1}.
\end{equation}
We then have:
\begin{equation}
e^{\alpha \hat T} \hat{\Delta} (\hat A -a) = \frac{1}{2 \pi} \int dk \hspace{0.2 cm} e^{\alpha \hat T} \cdot e^{i k (\hat A -a)} = \frac{1}{2 \pi} \int dk \hspace{0.2 cm} e^{ik (\hat A -a) + \mu \alpha \hat T}.
\end{equation}
From eqs.(28,33), we obtain upon application of the Weyl-map:
\begin{equation}
\begin{array}{c}
\Delta_* (A, a + n \lambda , a) = \left(T* \right)^n \Delta_* (A-a) = \frac{\partial^n}{\partial \alpha ^n} \left[ e_*^{\alpha T} * \Delta_* (A-a) \right]_{\alpha =0}=\\
\\
= \frac{(-1)^n}{2\pi} \int d \alpha \int dk \hspace{0.2 cm} \delta^{(n)} ( \alpha) e_*^{ik (A-a) + \mu \alpha  T}.
\end{array}
\end{equation}
Notice that the previous formula is equally valid if the spectrum is continuous. We assume that the finite "jump" $\lambda = b-a$ takes place in $N$ infinitesimal uniform steps of "length" $\epsilon$: $a_{i+1} - a_i = \epsilon$, $i=0,1, \cdots ,N$, with $a_0 =a$, $a_N =b $ and $\epsilon = \frac{b-a}{N}$. In that case, we have:
\begin{equation}
\hat T = \exp \left( - \frac{i}{\hbar} \epsilon \hat B \right) \simeq 1 - \frac{i}{\hbar} \epsilon \hat B.
\end{equation}
From the operator analog of eq.(34), we then have:
\begin{equation}
\begin{array}{c}
\left. \hat{\Delta} \left( \hat A, b , a \right) = \lim_{N \to + \infty} \frac{1}{2 \pi} \int dk \hspace{0.2 cm} \frac{\partial^N}{\partial \alpha^N} e^{ik (\hat A -a ) + \mu \alpha \hat T} \right|_{\alpha =0}= \left. \lim_{N \to + \infty} \frac{1}{2 \pi} \frac{\partial^N}{\partial \alpha^N} e^{\alpha \hat T } \int dk \hspace{0.2 cm}  e^{ik (\hat A -a )}  \right|_{\alpha =0} = \\
\\
= \lim_{N \to + \infty} \frac{1}{2 \pi} \left( 1- \frac{i}{\hbar} \frac{(b-a)}{N} \hat B \right)^N  \int dk \hspace{0.2 cm}  e^{ik (\hat A -a )} = \frac{1}{2 \pi} e^{- \frac{i}{\hbar} (b-a) \hat B} \int dk \hspace{0.2 cm}  e^{ik (\hat A -a )} = \\
\\
= \frac{1}{2 \pi}  \int dk \hspace{0.2 cm}  e^{- \frac{i}{\hbar} (b-a) \hat B + ik \left( \hat A - \frac{a + b}{2} \right)},
\end{array}
\end{equation}
where we used the fact that:
\begin{equation}
\lim_{N \to + \infty} \left( 1 + \frac{\hat X}{N} \right)^N = e^{\hat X}.
\end{equation}

\subsection{Basic properties of the stargenfunctions}

In this section we summarize several properties of the stargenfunctions:

1) Let $A(q,p)$ be a real symbol with non-degenerate spectrum. Then both the observables and the Wigner function can be expanded in terms of the functionals ${\Delta}_* (A,b,a)$. For the Wigner function the expansion reads:
\begin{equation}
F^W(q,p)= \int da db \left( \int dq'dp' F^W (q',p') \Delta_*(A(q',p'),a,b) \right) \Delta_*(A(q,p),b,a),
\end{equation}
and equally for a general observable:
\begin{equation}
X(q,p)= \int da db \left( \int dq'dp' X(q',p') \Delta_*(A(q',p'),a,b) \right) \Delta_*(A(q,p),b,a).
\end{equation}
Furthermore, if $X(q,p)=A(q,p)$ then eq.(39) reduces to:
$$
A(q,p)= \int da \hspace{0.2 cm} a \Delta_*(A(q,p)-a),
$$
this being the inverse formula of eq.(17).

2) The probability functional for a general observable is given by:
$$
P(A(q,p)=a) = \int dq dp F^W(q,p) \Delta_*(A(q,p)-a),
$$
and fully copies the analogous object of classical statistical mechanics. Moreover, the distribution $\Delta_*(A(q,p)-a)$ satisfies:
\begin{equation}
\frac{\int dq dp A(q,p) \Delta_*( A(q,p)-a)}{\int dq dp \Delta_*( A(q,p)-a)} = a.
\end{equation}
for all $a$ belonging to the spectrum of $A(q,p)$.
These two properties suggest that the probability of finding the observable $A$ with the value $a$ is given (just like in classical statistical mechanics) by the integration of the Wigner distribution function $F^W$ over the phase space hypersurface $A(q,p)=a$. In fact one has to be more careful: the distribution $\Delta_*(A-a)$ does not in general identify the hypersurface $A=a$, due to the non-local nature of the star product: $\Delta_*(A-a)$ is a "star delta function" which in general also assumes non zero values in phase space points not belonging to the hypersurface $A=a$. What can be said is that the star delta function $\Delta_*(A-a)$ identifies the hypersurface $A=a$ in the star phase space or, in other words, that it identifies the {\it star hypersurface} $A=a$.

3) We now prove that $\Delta_*(A)$ is a $\hbar$-deformation of the Dirac delta function. From eq.(16) and the definition of the star product it follows that:
\begin{equation}
e_*^{ikA} = \sum_{n=0}^{\infty} \frac{(ik)^n}{n!} \sum_{m_1,..,m_{n-1}=0}^{\infty} \frac{(i\hbar /2)^{m_1+..+m_{n-1}}}{m_1!..m_{n-1}!} A J^{m_1}A...AJ^{m_{n-1}}A,
\end{equation}
where $ J= \left(\frac{ {\buildrel { \leftarrow}\over\partial}}{\partial q} \frac{ {\buildrel {
\rightarrow}\over\partial}}{\partial p} -  \frac{{\buildrel { \leftarrow}\over\partial}}{\partial p}  \frac{{\buildrel {
\rightarrow}\over\partial}}{\partial q} \right) $. Let now $m_1+...+m_{n-1} =s$. From eqs.(17,41) it is clear that $\Delta_*(A)$ can be cast as a power series in $\hbar$:
\begin{equation}
\Delta_* (A) = \sum_{s=0}^{\infty} \left(\frac{i\hbar}{2}\right)^s \int dk \sum_{n=0}^{\infty}
\frac{(ik)^n}{n!}
\sum_{m_1+..+m_{n-1}=s} \frac{1}{m_1!..m_{n-1}!} A J^{m_1}A...AJ^{m_{n-1}}A,
\end{equation}
and is also trivial to check that the zero order term $(s=0)$ of the previous expression is  just $\delta(A)$. Hence, we conclude that $\Delta_*(A)$ is an $\hbar $-deformation of $\delta(A)$.

4) Let us calculate the explicit form of the former expansion up to the third order in $\hbar$. From the definition of $A^{*n}$ (eq.(16)) we get:
\begin{eqnarray}
A^{*n} &=& \sum_{s=0}^{\infty} \left(\frac{i\hbar}{2}\right)^s
\sum_{m_1+..+m_{n-1}=s} \frac{1}{m_1!..m_{n-1}!} A J^{m_1}A...AJ^{m_{n-1}}A \nonumber \\
&= &
A^n + \frac{1}{2} \left(\frac{i\hbar}{2}\right)^2 \Theta_1 n(n-1)A^{n-2} +
\frac{1}{6} \left(\frac{i\hbar}{2}\right)^2 \Theta_2 n(n-1)(n-2)A^{n-3} + {\cal O} (\hbar ^4),
\end{eqnarray}
where we used the notation: $\Theta_1= \frac{1}{2} AJ^2A = \frac{\partial^2A}{\partial q^2}\frac{\partial^2A}{\partial p^2}- \left(\frac{\partial^2A}{\partial q \partial p}\right)^2 $ and $ \Theta_2= \frac{1}{2} \{AJ^2A^2 -2A(AJ^2A)\}= \frac{\partial^2A}{\partial q^2} \left(\frac{\partial A}{\partial p}\right)^2
- 2\frac{\partial^2A}{\partial q \partial p} \frac{\partial A}{\partial q} \frac{\partial A}{\partial p} + \frac{\partial^2A}{\partial p^2} \left(\frac{\partial A}{\partial q}\right)^2$. It follows that:
\begin{equation}
e_*^{ikA} = \left[ 1+
\frac{1}{2} \left(\frac{i\hbar}{2}\right)^2 (ik)^2 \Theta_1 +
\frac{1}{6} \left(\frac{i\hbar}{2}\right)^2 (ik)^3 \Theta_2 \right] e^{ikA} + {\cal O} (\hbar ^4),
\end{equation}
and therefore:
\begin{equation}
\Delta_*(A) = \delta (A) - \frac{\hbar^2}{8} \Theta_1 \delta''(A) -
\frac{\hbar^2}{24} \Theta_2 \delta'''(A) + {\cal O} (\hbar ^4),
\end{equation}
Both eq.(42) and eq.(45) are in perfect agreement with previous derivations of the semiclassical form of a generic stargenfunction \cite{Balazs}.
It is however important to remark that these semiclassical expansions should be taken with caution. The expansion is useful if it truncates at some order. If, however, this is not the case then the expansion may not converge order-by-order. There are few known cases which can be dealt with appropriately (see for instance \cite{Berry,Almeida1,Almeida3}).

\section{Wigner quantum mechanics in the Heisenberg picture.}

The results of the previous sections lead to a complete formal solution of Wigner quantum mechanics in the Heisenberg picture. In this scheme the time evolution of a general observable $A(q,p)$ is given by the equation of motion:
\begin{equation}
\frac{\partial }{\partial t} A(q,p;t) = [A(q,p;t),H(q,p)]_M,
\end{equation}
which displays the formal solution:
\begin{equation}
A(q,p;t)=\sum_{n=0}^{+\infty}  \frac{t^n}{n!} [...[A(q,p;0),H(q,p)]_M...,H(q,p)]_M= U(t)^{-1} *A(q,p;0)* U(t),
\end{equation}
where $U(t)= e_*^{-itH(q,p)/\hbar}$ is the time propagator. Moreover,
the general stargenfunction of $A(q,p;t)$ is given by:
\begin{equation}
 \Delta_*(A(q,p;t)-a) =  \frac{1}{2\pi} \int dk e_*^{ik(A(q,p;t)-a)} =  U(t)^{-1} * \Delta_*(A(q,p;0)-a)* U(t) ,
\end{equation}
and thus it equally satisfies the time evolution equation:
\begin{equation}
\frac{\partial}{\partial t} \Delta_*(A(q,p;t)-a) =
[\Delta_*(A(q,p;t)-a),H(q,p)]_M.
\end{equation}
The previous equation together with the relation
$$
A(q,p;t)= \int da \hspace{0.2 cm} a \Delta_*(A(q,p;t)-a)
$$
lead to the conclusion that the stargenfunctions $\Delta_*(A(q,p;t)-a)$ encapsulate the entire information concerning the time evolution of the system.
In particular, the probability that a measurement of $A(q,p;t)$ at time $t$ yield the value $a$ is given by:
\begin{equation}
P(A(q,p;t)=a) = \int dq dp F^W(q,p) \Delta_*(A(q,p;t)-a),
\end{equation}
and satisfies the following suggestive formula:
$$
P(A(q,p;t)=a) =  < \Delta_*(A(q,p;t)-a) >.
$$
Consequently:
\begin{eqnarray}
\frac{\partial}{\partial t} P(A(q,p;t)=a) & = &  \int dq dp F^W(q,p)
[\Delta_*(A(q,p;t)-a),H(q,p)]_M =  \nonumber \\
 &=&  < [\Delta_*(A(q,p;t)-a),H(q,p)]_M >.
\end{eqnarray}
Finally, notice that the probability predictions can be easily connected with the corresponding formula in the Schr\"{o}dinger picture if one notices that:
\begin{eqnarray}
& & \int dq dp F^W(q,p) \left( U(t)^{-1}*\Delta_*(A(q,p;0)-a)*U(t)\right) \nonumber \\
&=& \int dq dp \left( U(t)* F^W(q,p) * U(t)^{-1}\right) \Delta_*(A(q,p;0)-a).
\end{eqnarray}

\section{Solution for quadratic phase space functionals}

In this section we will derive a general expression for the stargenfunctions associated with the Weyl symbol of a generic quadratic operator. We will address the particular case when the Hessian matrix of such an operator is proportional to a symplectic matrix. In this instance we will analyze thoroughly continuous and discrete spectra. All these formulae are applicable both to diagonal and non-diagonal stargenfunctions and they generalize the results of ref.\cite{Flato1}.

The calculation of the $*$-exponential is the most crucial step. Following a technique developed in ref.\cite{Flato1}, we can derive the $*$-exponential for any polynomial of degree 2 in arbitrary dimension. To be more precise, let:
\begin{equation}
{\cal A} = z^T A z+ b^T z,
\end{equation}
where $A$ is a symmetric, non-singular, $2N \times 2N$ matrix, the superscript $T$ stands for matrix transposition, and:
\begin{equation}
z^T = \left( p_1, \cdots, p_N , q_1, \cdots , q_N  \right) , \hspace{0.5 cm} b^T = \left( b_1, \cdots, b_N , b_{N+1}, \cdots , b_{2N}  \right).
\end{equation}
The restriction that the matrix $A$ be non-singular is not indispensable as we shall see in the example of the linear potential. We then prove the following theorem.

\vspace{0.5 cm}
\noindent
{\underline{\bf Theorem:}} The non-commutative exponential is given by:
\begin{equation}
e_*^{\beta {\cal A}} = \left( \mbox{det} \cos B \right)^{-1 /2} \exp \left\{ - \frac{1}{\hbar} \left(z^T + \frac{1}{2} b^T A^{-1} \right) J S_A^{-1} J \tan B J S_A  \left(z + \frac{1}{2}  A^{-1} b \right) - \frac{\beta}{4} b^T A^{-1} b \right\},
\end{equation}
where
\begin{equation}
B = \hbar \beta J S_A J S_A^T J,
\end{equation}
$J$ is the $2N \times 2 N$ symplectic form:
\begin{equation}
J = \left(
\begin{array}{c r}
0 & -1\\
1 & 0
\end{array}
\right)
\end{equation}
and $S_A$ is a $2N \times 2N$ matrix which satisfies:
\begin{equation}
S_A^T S_A =A.
\end{equation}
The corresponding $*$-genfunction then reads:
$$
\begin{array}{c}
\Delta_* (A - a) = \frac{1}{2\pi} \int d k \hspace{0.2 cm} \left[ \mbox{det} \cos B (k) \right]^{-1 /2} \times \\
\\
\times \exp \left\{ - \frac{1}{\hbar} \left(z^T + \frac{1}{2} b^T A^{-1} \right) J S_A^{-1} J \tan B (k) J S_A  \left(z + \frac{1}{2}  A^{-1} b \right) - ik \left(a+ \frac{1}{4} b^T A^{-1}b \right)  \right\},
\end{array}
$$
where $B(k)$ is given by (56) with $\beta$ replaced by $ik$.

\vspace{0.5 cm}
\noindent
{\underline{\bf Proof:}} Let us first calculate $e_*^{\beta {\cal D}}$ with ${\cal D} = z^T A z$. According to (16) we have:
\begin{equation}
e_*^{\beta {\cal D}} = \sum_{n=0}^{+ \infty} \frac{\beta^n}{n!} \Omega_n,
\end{equation}
where:
\begin{equation}
\Omega_{n+1} = {\cal D} * \Omega_n = \left[{\cal D} + \frac{\hbar^2}{4} (JAJ)_{ij} \partial_i \partial_j \right] \Omega_n, \hspace{0.5 cm} \Omega_0 =1.
\end{equation}
In this expression, sum over repeated indices is understood and $\left\{\partial_i \right\} = \left( \frac{\partial}{\partial p_1}, \cdots , \frac{\partial}{\partial p_N}, \frac{\partial}{\partial q_1} , \cdots, \frac{\partial}{\partial q_N} \right)$. It is easy to check that the imaginary terms in the $*$-product (i.e. terms proportional to odd powers of $\hbar$) do not contribute to eq.(60).

Let us now consider a function $\phi \left( \beta, z \right)$ which satisfies:
\begin{equation}
\frac{\partial \phi}{\partial \beta} = {\cal D} \phi + \frac{\hbar^2}{4} \left( J A J \right)_{ij} \partial_i \partial_j \phi, \hspace{0.5 cm} \phi(0,z) = 1.
\end{equation}
If we expand $\phi \left( \beta, z \right)$ in powers of $\beta$, $\phi \left(\beta , z \right) = \sum_{n=0}^{+ \infty} \frac{\beta^n}{n!} K_n (z)$, and substitute in eq.(61), we then get the following recursive relations:
\begin{equation}
K_{n+1} (z) = \left[{\cal D} + \frac{\hbar^2}{4} (JAJ)_{ij} \partial_i \partial_j \right] K_n (z), \hspace{0.5 cm} K_0 (z) =1.
\end{equation}
Upon comparison with (60), we conclude that $\Omega_n (z) = K_n (z)$, $\forall z$ and hence: $\phi \left( \beta , z \right) = e_*^{\beta {\cal D}}$. To obtain the non-commutative exponential, we thus have to solve (61). We consider the following Ansatz:
\begin{equation}
\phi \left( \beta ,z \right) = M \left( \beta \right) \exp \left[ z^T \Lambda \left( \beta \right) z \right],
\end{equation}
where $M \left( \beta \right)$ is a scalar function and $\Lambda \left( \beta \right) $ is a symmetric $2N \times 2 N$ matrix-valued function. If we substitute (63) in (61), we get:
\begin{equation}
\left\{
\begin{array}{l}
\frac{M'}{M} = \frac{\hbar^2}{2} Tr \left( JAJ \Lambda \right),\\
\Lambda' = A + \hbar^2 \Lambda J A J \Lambda ,
\end{array}
\right.
\end{equation}
where $M' \left( \beta \right)= \frac{d M}{d \beta}$ and $\Lambda' \left( \beta \right)= \frac{d \Lambda}{d \beta}$. Notice that:
\begin{equation}
J^{-1} = J^T = - J.
\end{equation}
If we define the matrix $B$ as in (56), it then follows that $B$ and $\tan B$ are antisymmetric matrices. Likewise we define the matrix $\Lambda $ as:
\begin{equation}
\Lambda = - \frac{1}{\hbar} J S_A^{-1} J \tan B J S_A.
\end{equation}
This matrix is symmetric. Indeed, we have from (65):
\begin{equation}
\Lambda^T = - \frac{1}{\hbar} S_A^T J \tan B J \left( S_A^{-1} \right)^T J = \frac{1}{\hbar} J S_A^{-1} J \left[J S_A J S_A^T J \tan B J \left( S_A^{-1} \right)^T J S_A^{-1} J \right] J S_A.
\end{equation}
However, since $B = \hbar \beta J S_A J S_A^T J$, we conclude that:
\begin{equation}
J S_A J S_A^T J \tan B J \left( S_A^{-1} \right)^T J S_A^{-1} J = \tan B J S_A J S_A^T J^2 \left( S_A^{-1} \right)^T J S_A^{-1} J = - \tan B.
\end{equation}
Eq.(67) then reads:
\begin{equation}
\Lambda^T = - \frac{1}{\hbar} J S_A^{-1} J \tan B J S_A = \Lambda.
\end{equation}
From (66) we have:
\begin{equation}
\Lambda' = - J S_A^{-1} J^2 S_A J S_A^T J \left( \cos^2 B \right)^{-1} J S_A = - S_A^T J \left( 1 + \tan^2 B \right) J S_A  = A - S_A^T J \tan^2 B J S_A.
\end{equation}
On the other hand we have:
\begin{equation}
\hbar^2 \Lambda^T J A J \Lambda = S_A^T J \tan B J \left( S_A^{-1} \right)^T J^2 S_A^T S_A J^2 S_A^{-1} J \tan B J S_A = - S_A^T J \tan^2 B J S_A,
\end{equation}
and we conclude that (66) is indeed a solution of (64). It remains to prove that:
\begin{equation}
M \left( \beta \right) = \left( \det \cos B \right)^{-1/2} .
\end{equation}
For a generic matrix $G$, we have:
\begin{equation}
\frac{d}{d \beta} \det G \left( \beta \right) = \det G \left( \beta \right) Tr \left[G^{-1} \left( \beta \right) \frac{d}{d \beta} G \left( \beta \right) \right].
\end{equation}
It then follows that:
\begin{equation}
\begin{array}{c}
\frac{d}{d \beta} M \left( \beta \right) = \frac{d}{d \beta} \left( \det \cos B \left( \beta \right) \right)^{-1 /2} = - \frac{1}{2} M^3 \left( \det \cos B \right) Tr \left[ \left( \cos B \right)^{-1} \frac{d}{d \beta } \cos B \left( \beta \right) \right] =\\
\\
=  \frac{\hbar}{2} M Tr \left[\tan B J S_A J S_A^T J \right].
\end{array}
\end{equation}
Moreover we have:
\begin{equation}
\frac{\hbar^2}{2} Tr \left( J A J \Lambda \right) = - \frac{\hbar}{2} Tr \left(J A J^2 S_A^{-1} J \tan B J S_A \right) = \frac{\hbar}{2} Tr \left( \tan B J S_A J S_A^T J \right),
\end{equation}
where we used the cyclicity of the trace. Comparing (74) and (75), we conclude that (72) is indeed a solution of (64).

Finally, we prove eq.(55). Let us define $v = z + \frac{1}{2} A^{-1} b $. We then have: ${\cal D} (v) = v^T A v = {\cal A} (z) + \frac{1}{4} b^T A^{-1} B$. Notice that the translation $z \longrightarrow v$ does not affect the $*$-product. Consequently:
\begin{equation}
e_{*(z)}^{\beta {\cal A} (z)} =  e_{*(z)}^{\beta {\cal D} (v) - \frac{\beta}{4} b^T A^{-1} b} = e^{- \frac{\beta}{4} b^T A^{-1} b} e_{*(v)}^{\beta {\cal D} (v)} = \left( \det \cos B \right)^{- 1/2} \exp \left[ v^T \Lambda v - \frac{\beta}{4} b^T A^{-1} b \right ],
\end{equation}
where we have used the notation $*(z)$, $*(v)$ to specify the variables with respect to which we compute the $*$-products. From the expressions for $\Lambda$ and $v$, we recover eq.(55).$_{\Box}$

\subsection{Determining the matrix $S_A$}

A few comments are in order concerning the existence and uniqueness of the matrix $S_A$ defined by eq.(58). First of all, notice that $S_A$ is not unique. In fact it is defined modulo some orthogonal matrix: if $O$ is a $2N \times 2N$ matrix such that $OO^T= O^T O =I$, then if $A= S_A^T S_A$, we also have $A= S_A'^T S_A'$ where $S_A' = O S_A$. On the other hand, any quadratic form $Q(z)$ can be algebraically diagonalized, which means that if $Q(z) = \sum_{1 \le i \le j \le 2N} a_{ij} z_i z_j$, then there exists a set of variables $v_i$ such that $Q(z) = Q' (v)= \sum_{i=1}^{2N} \lambda_i v_i^2$, where $\left\{ \lambda_i, \hspace{0.3 cm} i =1, \cdots , 2N \right\}$ is a set of complex numbers (possibly, not all distinct). This can be expressed in matrix language as follows:
$$
v = P^{-1} z, \hspace{0.5 cm} Q(z) = z^T A z = v^T D v = Q' (v),
$$
where:
$$
D= diag \left( \lambda_1 , \cdots, \lambda_{2N} \right) = P^T A P.
$$
Notice that in the previous expressions, in general $P^T \ne P^{-1}$ and $P$ is a complex matrix. The exception is when $A$ is real, in which case $P$ is a real orthogonal matrix.

To proceed we define the matrix $D^{\frac{1}{2}} = diag \left( \sqrt{\lambda_1}, \cdots, \sqrt{\lambda_{2N}} \right)$. The matrix $S_A$ then reads:
$$
S_A = D^{\frac{1}{2}} P^{-1}.
$$
In some cases, when diagonalizing the matrix in this fashion is too complicated, we can alternatively determine $S_A$ by a straightforward generalization of Cholesky's method to complex matrices. In that case, we look for a decomposition of the form $S_A= U$, where $U$ is an upper triangular complex matrix. Its elements can be computed iteratively according to:
$$
\left\{
\begin{array}{l}
u_{ii} = \left( a_{ii} - \sum_{k=1}^{i-1} u_{ki}^2 \right)^{\frac{1}{2}}, \\
\\
u_{ii} u_{ij} = a_{ij} - \sum_{k=1}^{j-1} u_{ki} u_{kj},
\end{array}
\right.
i=1, \cdots , 2N, \hspace{0.3 cm} j = i+1, \cdots, 2N.
$$
To specify when this method is applicable to a complex matrix $A$ let us define the set of $2N$ matrices $A_i$ $(i=1, \cdots , 2N)$:
$$
A_1 = \left(a_{11} \right), \hspace{0.5 cm} , A_2 = \left(
\begin{array}{c c}
a_{11} & a_{12}\\
a_{12} & a_{22}
\end{array}
\right), \hspace{0.5 cm} A_3 = \left(
\begin{array}{c c c}
a_{11} & a_{12} & a_{13}\\
a_{12} & a_{22} & a_{23}\\
a_{13} & a_{23} & a_{33}
\end{array}
\right), \cdots
$$
The method is then valid whenever $\det (A_i) \ne 0$, $\forall i =1, \cdots , 2N$.

\subsection{A particular case}

The formula (55) can be drastically simplified in the following particular case. Suppose the matrix $S_A$ is proportional to a symplectic matrix, i.e.:
\begin{equation}
S_A J S_A^T = \alpha J,
\end{equation}
where $\alpha$ is some nonzero constant. If a matrix is symplectic, then so is its transpose. It then follows that:
\begin{equation}
S_A^T J S_A = \alpha J.
\end{equation}
Since $S_A^T S_A = A$, we get:
\begin{equation}
A J A = S_A^T S_A J S_A^T S_A = \alpha S_A^T J S_A = \alpha^2 J.
\end{equation}
Moreover,
\begin{equation}
S_A J S_A^T = \alpha J \Longrightarrow  S_A J S_A^T J = -\alpha I,
\end{equation}
where $I$ is the $2N \times 2 N$ identity matrix. Consequently,
\begin{equation}
S_A^{-1} = - \frac{1}{\alpha} J S_A^T J, \hspace{0.5 cm} A^{-1} = - \frac{1}{\alpha^2} J A J.
\end{equation}
From eq.(56), we have:
\begin{equation}
B = \hbar \beta J S_A J S_A^T J = - \hbar \alpha \beta J.
\end{equation}
And so:
\begin{equation}
\left\{
\begin{array}{l}
B^{2n} = \left( i \hbar \alpha \beta \right)^{2n} I\\
B^{2n +1} = i \left( i \hbar \alpha \beta \right)^{2n +1} J,
\end{array}
\right.
\end{equation}
from which follows:
\begin{equation}
\sin B = i \sin \left( i \hbar \alpha \beta \right) J, \hspace{0.5 cm} \cos B =  \cos \left( i \hbar \alpha \beta \right) I , \hspace{0.5 cm} \tan B = i \tan \left( i \hbar \alpha \beta \right) J.
\end{equation}
We then have:
\begin{equation}
\Lambda = - \frac{1}{\hbar} J S_A^{-1} J \tan B J S_A = \frac{i}{\alpha \hbar} \tan \left( i \hbar \alpha \beta \right) J^2 S_A^T J^4 S_A = - \frac{i}{ \alpha \hbar} \tan \left( i \hbar \alpha \beta \right) A,
\end{equation}
and
\begin{equation}
M = \left( \det \cos B \right)^{- \ 1/2} = \left[\det \cos \left( i \hbar \alpha \beta \right) I \right]^{- 1/2} = \left[\cos \left( i \hbar \alpha \beta \right) \right]^{-N}.
\end{equation}
Equation (55) thus reads:
\begin{equation}
e_*^{\beta {\cal A}} = \left[\cos \left( i \hbar \alpha \beta \right) \right]^{-N} \exp \left\{\frac{{\cal A}}{\alpha i \hbar} \tan \left( i \hbar \alpha \beta \right) + \frac{1}{4 \alpha i \hbar} b^T A^{-1} b \left[ \tan \left( i \hbar \alpha \beta \right) -  i \hbar \alpha \beta \right] \right\},
\end{equation}
where $A^{-1}$ is given by (81).

How interesting is this specific case? To begin with, it applies to the harmonic oscillator in arbitrary dimension. Furthermore it is valid for any one-dimensional system. Indeed, a $2 \times 2$ symplectic matrix is any matrix of determinant $1$. So any nonsingular matrix is proportional to a symplectic matrix. On the other hand, even if the $2 \times 2$ matrix is singular (which entails $\alpha =0$), we can still apply the previous results if we replace the vanishing eigenvalue by some arbitrary infinitesimal parameter. This will be our approach in the case of the linear potential (cf. section 6).

Let us then consider some examples. The ensuing analysis will slightly generalize the results of ref.\cite{Flato1}. We assume that eqs.(77,78) hold for $\alpha \ne 0$ and that $b=0$. We now prove the following lemma:

\vspace{0.3 cm}
\noindent
{\underline{\bf Lemma:}} If $A=S_A^T S_A$ is hermitean, then $\alpha$ is either real or purely imaginary.

\vspace{0.3 cm}
\noindent
{\underline{\bf Proof:}} If A is hermitean then it follows from (81)\footnote{The superscripts $*$ and $\dagger$ denote complex conjugation and hermitian conjugation, respectively.}:
\begin{equation}
S_A^T S_A = S_A^{\dagger} S_A^* \Longrightarrow S_A^T = S_A^{\dagger} S_A^* S_A^{-1} = - \frac{1}{\alpha} S_A^{\dagger} S_A^* J S_A^T J.
\end{equation}
From this we infer that:
\begin{equation}
\left\{
\begin{array}{l}
S_A = - \frac{1}{\alpha} J S_A J S_A^{\dagger} S_A^*,\\
S_A^* = - \frac{1}{\alpha^*} J S_A^* J S_A^T S_A,\\
S_A^{\dagger} = - \frac{1}{\alpha^*} S_A^T S_A J S_A^{\dagger} J.
\end{array}
\right.
\end{equation}
If we hermitean conjugate eq.(77) and use the previous equation , we get:
\begin{equation}
\begin{array}{c}
S_A^* J S_A^{\dagger} = \alpha^* J \Longleftrightarrow \left(\frac{1}{\alpha^*} \right)^2 J S_A^* J S_A^T S_A J S_A^T S_A J S_A^{\dagger} J  = \alpha^* J\\
\Longleftrightarrow \frac{\alpha}{\left( \alpha^* \right)^2}  J S_A^* J S_A^T J S_A J S_A^{\dagger} J  = \alpha^* J \Longleftrightarrow \left(\frac{\alpha}{\alpha^*} \right)^2   J S_A^* J^3 S_A^{\dagger} J  = \alpha^* J\\
\Longleftrightarrow \frac{\alpha^2}{\alpha^*}  J  = \alpha^* J \Longleftrightarrow \alpha^2 = \left( \alpha^* \right)^2 \Longleftrightarrow \alpha = \pm \alpha^*._{\Box}
\end{array}
\end{equation}

If $\alpha =0$, then we have to regularize the matrix $A$ as we shall see in the example of the linear potential.

If $\alpha $ is real (and nonzero), then the $*$-exponential in eq.(87) yields for $\beta = i k$ \cite{Flato1}:
\begin{equation}
e_*^{ik  {\cal A}} = \left[\cos \left( \hbar \alpha k \right) \right]^{-N} \exp \left[\frac{i {\cal A}}{\alpha \hbar}  \tan \left( \hbar \alpha k  \right) \right].
\end{equation}
For fixed ${\cal A} \ne 0$, this can be viewed as an analytical distribution in the variable $k \in {\cal C} \backslash \left\{ (2n +1) \frac{\pi}{2 \hbar \alpha} , \hspace{0.3 cm} n \in {\cal Z} \right\}$. Being a periodic distribution (with period  $\frac{2 \pi}{\alpha \hbar}$), it can be expanded in a Fourier series:
\begin{equation}
\phi ( k) \equiv e_*^{i k {\cal A}} = \sum_{n = - \infty}^{+ \infty} C_n e^{ in \alpha \hbar k},
\end{equation}
where the Fourier modes are given by:
\begin{equation}
C_n = \frac{\alpha \hbar}{2 \pi} \int_{- \frac{\pi}{\alpha \hbar}}^{\frac{\pi}{\alpha \hbar}} d k \hspace{0.2 cm} \phi (k) e^{-in \alpha \hbar k }.
\end{equation}
If we make the substitution $z = e^{i \alpha \hbar k }$, we get:
\begin{equation}
C_n = \frac{2^N}{2 \pi i} \oint_C dz \hspace{0.2 cm} z^{-n + N} (z^2 +1)^{-N} \exp \left[ \frac{{\cal A}}{\alpha \hbar} \left( \frac{z^2 -1}{z^2 +1} \right) \right].
\end{equation}
Notice that the integral in eq.(93) is improper because of the singularities at $k= \pm  \frac{\pi}{2 \alpha \hbar}$. This can be dealt with, if we choose $C$ to be some contour in the complex $z$-plane containing the origin in its interior, but not the singularities $z= \pm i$. Cauchy's theorem shows that $C_n$ vanishes for $n \le N$ and for $n$ even. A straightforward calculation yields:
\begin{equation}
C_{2n+ N + 1} = 2^N (-1)^n e^{- \frac{{\cal A}}{\alpha \hbar}} L_n^{N-1} \left(\frac{2{\cal A}}{\alpha \hbar} \right), \hspace{0.5 cm} n=0, 1,2, \cdots,
\end{equation}
where $L_n^{\alpha} (x)$ are the Laguerre polynomials. This can be proven by using the following representation of $L_n^{\alpha} (x) \cite{Grad}$:
\begin{equation}
L_n^{\alpha} (x) = \sum_{k=0}^n (-1)^k \left(
\begin{array}{c}
n + \alpha\\
n-k
\end{array}
\right) \frac{x^k}{k!}.
\end{equation}
An interesting fact of this approach is that, as a by-product, we obtain an integral representation for the Laguerre polynomials. Indeed if we equate eqs.(94) and (95), we get:
\begin{equation}
L_n^{\alpha} (x) = \frac{(-1)^n}{2 \pi i} \oint_C dz \hspace{0.2 cm} z^{-2n-1} (z^2+1)^{- \alpha - 1}  \exp \left( \frac{x z^2}{z^2 +1} \right), \hspace{0.5 cm} n , \alpha  =0,1,2 , \cdots
\end{equation}
A remark concerning the normalization is now in order. In the previous equations the coefficients $C_{2n+N+1}$ are not the normalized Wigner functions. In fact we have:
\begin{equation}
\int d^N q \int d^N p \hspace{0.2 cm} C_{2n+N+1} (q,p)= \left(2 \pi \hbar \right)^N , \hspace{0.5 cm} n =0,1,2 , \cdots
\end{equation}
Note that this is consistent with the fact that $C_{2n+N+1} = W \left( |n><n| \right) $ and that the Wigner function is $F_{nn}^W = \frac{1}{\left(2 \pi \hbar\right)^N} W \left( |n><n| \right) $. Consequently:
\begin{equation}
F_{nn}^W = \frac{(-1)^n}{(\pi \hbar)^N} e^{- \frac{{\cal A}}{\alpha \hbar}} L_n^{N-1} \left( \frac{2 {\cal A}}{\alpha \hbar} \right), n= 0 ,1,2 \cdots
\end{equation}
From the previous analysis, eq.(17) reads:
\begin{equation}
\Delta_* \left( {\cal A} - a \right) = \frac{1}{2 \pi} \int dk \hspace{0.2 cm} \phi (k) e^{-ika}= \sum_{n=0}^{+ \infty} C_{2n+N+1} \delta \left(a - a_n \right), \hspace{0.5 cm} a_n = \hbar \alpha (2n+1).
\end{equation}

If, on the other hand, $\alpha = i\gamma$ $(\gamma \in \Re \setminus \left\{ 0 \right\})$, we have:
\begin{equation}
\begin{array}{c}
\Delta_* \left( {\cal A} - a \right) = \frac{1}{2 \pi} \int dk \hspace{0.2 cm} \left[\cosh \left( \hbar \gamma k \right) \right]^{- N} \exp \left[ \frac{\cal A}{\alpha \hbar} \tanh \left(\hbar \gamma k \right) - i k a \right] =\\
\\
= \frac{2^{N-2}}{\gamma \pi \hbar} \frac{\Gamma \left( \frac{N}{2} + \frac{ia }{2 \hbar \gamma} \right) \Gamma \left( \frac{N}{2} - \frac{ia }{2 \hbar \gamma} \right)}{\Gamma (N) } e^{\frac{\cal A}{i \hbar \gamma}}  \hspace{0.2 cm}_1F_1 \left( \frac{N}{2} + \frac{ia }{2 \hbar \gamma} ; N ; \frac{2 i {\cal A}}{\hbar \gamma} \right),
\end{array}
\end{equation}
where $_1F_1$ is the confluent hypergeometric function \cite{Grad}.

These results are a slight generalization of those in ref\cite{Flato1}, where Bayen et al. derived the expressions for continuous and discrete spectra, when the matrix $A$ takes the form:
$$
A=\left(
\begin{array}{c c}
a I_{2 \times 2} & b I_{2 \times 2}\\
b I_{2 \times 2} & c I_{2 \times 2}
\end{array}
\right),
$$
for arbitrary real constants $a,b,c$. This corresponds to $\alpha = \sqrt{ac - b^2}$ in eq.(77), which may be real or purely imaginary. Obviously eq.(77) admits more solutions for $A$ than the previous ones.

\section{The one-dimensional harmonic oscillator}

The aim of this section is to determine the diagonal and non-diagonal Wigner functions from the formal solution of the energy stargenvalue equation (3) for the simple harmonic oscillator.

The Hamiltonian reads $\hat H= \frac{1}{2} (\hat p^2 + \hat q^2)$, where to make it simpler we made $m=\hbar =w=1$. Consequently:
\begin{equation}
A= \frac{1}{2} I_{2 \times 2} , \hspace{0.5 cm} b^T = \left( 0 ,0 \right), \hspace{0.5 cm} AJA = \frac{1}{4} J.
\end{equation}
Since $\alpha = \frac{1}{2}$, it follows immediately from (99,100) that:
\begin{equation}
\Delta_* (H - E) = 2 \pi \sum_{n=0}^{+ \infty} F_{nn}^W (q,p) \delta \left( E - E_n \right),
\end{equation}
where:
\begin{equation}
F_{nn}^W (q,p) = \frac{(-1)^n}{\pi} e^{-2 H} L_n (4 H), \hspace{0.5 cm} E_n = n + \frac{1}{2}.
\end{equation}

Let us now derive an expression for the non-diagonal elements $F_{nm}^W (q,p)$. The creation and annihilation operators are defined by:
\begin{equation}
\hat a = \frac{\hat q + i \hat p}{\sqrt 2}, \hspace{0.5 cm} \hat a^{\dagger} = \frac{\hat q - i \hat p}{\sqrt 2} , \hspace{1.0 cm} \left[ \hat a , \hat a^{\dagger} \right]=1.
\end{equation}
Setting $E'= E + n$, eq.(34) reads:
\begin{equation}
\Delta_* (H,E',E) =  W \left( |E'>< E| \right) = \frac{(-1)^n}{2 \pi} \int d \alpha \int d k \hspace{0.2 cm} \delta^{(n)} (\alpha) e_*^{ik (H-E) + \alpha \mu a^{\dagger}}.
\end{equation}
The evaluation of the $*$-exponential yields:
\begin{equation}
e_*^{ik H + \alpha \mu a^{\dagger}} = e^{\alpha a^{\dagger}} \varphi \left( q - \frac{\alpha}{2 \sqrt 2} , p + \frac{i \alpha}{2 \sqrt 2} \right),
\end{equation}
where $\varphi (q,p) = e_*^{ik H}$. Consequently,
\begin{equation}
\Delta_* (H, E',E) = \frac{1}{2 \pi} \int d k \hspace{0.2 cm} e^{-ik E} \psi^{(n)} (k),
\end{equation}
where $\psi^{(n)} (k) $ is an analytical periodic distribution (with period $4 \pi$) in the variable $k \in {\cal C} \backslash \left\{(2n+1) \pi, \hspace{0.3 cm} n \in {\cal Z} \right\}$:
\begin{equation}
\psi^{(n)} (k) = \frac{\partial^n}{\partial \alpha^n} \left[ \frac{e^{\alpha a^{\dagger} \left(1 - i \tan (k/2) \right)}}{\cos (k/2)} e^{2i H \tan (k/2)} \right]_{\alpha =0}.
\end{equation}
Its Fourier expansion is:
\begin{equation}
\psi^{(n)} (k) = \sum_{l=0}^{+ \infty} d_{-2l-1} e^{-i (2l+1) k/2},
\end{equation}
with:
\begin{equation}
\begin{array}{c}
d_{-2l-1} (q,p) = \frac{1}{i \pi} \oint_C dz \hspace{0.2 cm} z^{-2l-1} (z^2+1)^{-1} \left( \frac{2 a^{\dagger} }{z^2 +1} \right)^n e^{2H \left( \frac{z^2 -1}{z^2 +1} \right)}=\\
\\
= \left. \frac{\left(2 a^{\dagger} \right)^n}{i \pi} e^{-2 H} \frac{\partial^n}{\partial x^n} \oint_C dz \hspace{0.2 cm} z^{-2(l+n)-1} (z^2+1)^{-1} e^{\frac{x z^2}{z^2 +1}}
\right|_{x=4 H}= \left( 2 a^{\dagger} \right)^n e^{- 2H} 2 (-1)^l L_l^n (4H).
\end{array}
\end{equation}
Substituting (110) in eq.(108), we get:
\begin{equation}
\Delta_* (H, E', E) = \sum_{l=0}^{+ \infty} \delta \left( E - E_l \right) d_{-2l -1}.
\end{equation}
The normalized Wigner function $F_{ml}^W $, (with $m=n+l$ and $\beta_{m,l} = \sqrt{\frac{l!}{m!}}$) is given by:
\begin{equation}
F_{ml}^W (q,p) = \frac{\beta_{m,l}}{2\pi} d_{-2l -1} = \frac{e^{- 2H}}{\pi} (-1)^l \sqrt{\frac{l!}{m!}} \left( 2 a^{\dagger} \right)^{m-l} L_l^{m-l} (4H), \hspace{0.5 cm} (m \ge l \ge 0).
\end{equation}
The analogous result for $l \ge m$ is tantamount to performing the substitutions $l \longleftrightarrow  m$ and $a^{\dagger} \longleftrightarrow a $ in the previous expression. This can be summarized in the following formula:
\begin{equation}
\begin{array}{c}
F_{nm}^W (q,p) =  \frac{e^{- 2H}}{\pi} (-1)^{(n+m - |m-n|)/2} \left( \frac{n!}{m!} \right)^{\frac{1}{2} sign (m-n)}  \left( 2 a \right)^{\frac{|m-n| + m -n}{2}} \times \\
\\
\times \left( 2 a^{\dagger} \right)^{\frac{|m-n| + n -m}{2}} L_{\frac{n+m - |m-n|}{2}}^{|m-n|} (4H), \hspace{0.5 cm} n,m = 0,1,2, \cdots
\end{array}
\end{equation}
Notice that $\left( F_{nm}^W (q,p) \right)^* = F_{mn}^W (q,p)$, which is agreement with the quantum mechanical counterpart $\left( |n><m| \right)^{\dagger} = |m>< n|$. A lengthy calculation shows that $F_{nm}^W$ obeys the correct $*$-genvalue equations:
\begin{equation}
H*F_{nm}^W = E_n F_{nm}^W, \hspace{1.0 cm} F_{nm}^W *H = E_m F_{nm}^W.
\end{equation}

\section{The linear potential}

The hamiltonian for a particle coupled to a linear potential has a continuous spectrum. It is defined by:
\begin{equation}
\hat H (q,p) = \frac{\hat p^2}{2} + \hat q.
\end{equation}
We then have:
\begin{equation}
A = \left(
\begin{array}{c c}
\frac{1}{2} & 0\\
0 & 0
\end{array}
\right),
\hspace{0.5 cm} b^T = \left(0,1 \right)
\end{equation}
Notice that the matrix $A$ is singular. To solve this problem, let us define:
\begin{equation}
A_{\lambda} = \left(
\begin{array}{c c}
\frac{1}{2} & 0\\
0 & \lambda
\end{array}
\right),
\hspace{0.5 cm}
A_{\lambda}^{-1} = \left(
\begin{array}{c c}
2 & 0\\
0 & \frac{1}{\lambda}
\end{array}
\right).
\end{equation}
We shall compute the $*$-exponential for finite $\lambda$ and then eventually take the limit $\lambda \to 0$. From these expressions we get:
\begin{equation}
b^T A_{\lambda}^{-1} b = \frac{1}{\lambda}, \hspace{0.5 cm} A_{\lambda} J A_{\lambda} = \frac{\lambda}{2} J \Longrightarrow \alpha_{\lambda} = \sqrt{\frac{\lambda}{2}}.
\end{equation}
And so:
\begin{equation}
e_*^{i k H} = \lim_{\lambda \to 0} \left[\cos \left(\alpha_{\lambda} k \right) \right]^{-1} \exp \left\{\frac{i H}{\alpha_{\lambda}} \tan \left( \alpha_{\lambda} k \right) + \frac{i}{4 \lambda \alpha_ {\lambda} } \left[\tan \left( \alpha_{\lambda} k \right) - \alpha_{\lambda} k \right] \right\} = e^{ ik  H  + \frac{i k^3}{24}}.
\end{equation}
It then follows that:
\begin{equation}
\Delta_* (H-E) = \frac{1}{2 \pi} \int dk \hspace{0.2 cm} e^{ik (H-E) + \frac{i k^3}{24}} = 2 Ai \left[2 \left( H - E \right) \right] ,
\end{equation}
where $Ai (z)$ is the Airy function \cite{Grad}:
\begin{equation}
Ai (x) = \frac{1}{2 \pi} \int dk \hspace{0.2 cm} e^{-i k x - \frac{i}{3} k^3}.
\end{equation}
This is a well-known result \cite{Fairlie1}. To compute the non-diagonal elements, we need the operator $\hat B$ which implements the infinitesimal translations in the spectrum of $\hat H$. A straightforward calculation shows that $\hat B =  \hat p$. We then have to compute the $*$-exponential (87), with:
\begin{equation}
\beta = i k, \hspace{0.5 cm}  A_{\lambda} = \left(
\begin{array}{c c}
\frac{1}{2} & 0\\
0 & \lambda
\end{array}
\right),
\hspace{0.5 cm}
A_{\lambda}^{-1} = \left(
\begin{array}{c c}
2 & 0\\
0 & \frac{1}{\lambda}
\end{array}
\right),
\hspace{0.5 cm} b^T = \left(- \frac{i}{\beta} (E' - E), 1 \right).
\end{equation}
After a simple calculation eq.(15) reads:
\begin{equation}
\Delta_* (H,E',E) =  2 e^{- i (E'-E) p} Ai \left[2 \left( H - \frac{E+E'}{2} \right) \right].
\end{equation}
Notice that the previous expression reduces to (121) if $E'=E$. Moreover using the fact that $Ai(x)$ satisfies the equation, $\frac{d^2}{d x^2} Ai (x) = x Ai (x)$, we can prove that:
\begin{equation}
\left(\frac{p^2}{2} + q \right) * \Delta_* \left(H, E',E \right) = E' \Delta_* \left( H,E',E \right), \hspace{0.5 cm} \Delta_* \left(H, E',E \right) * \left(\frac{p^2}{2} + q \right) = E \Delta_* \left(H, E',E \right).
\end{equation}

\section{Two-dimensional harmonic oscillator}

The Hamiltonian for this model is:
\begin{equation}
\hat H = \frac{\hat p_1^2}{2} + \frac{\hat p_2^2}{2} + \frac{\hat q_1^2}{2} + \frac{\hat q_2^2}{2},
\end{equation}
Our purpose is to determine the simultaneous $*$-genfunctions of the hamiltonian and of the $z$ component of the angular momentum:
\begin{equation}
\hat L_3 = \hat q_1 \hat p_2 - \hat p_1 \hat q_2.
\end{equation}
Since the two observables commute our first step is to compute the $*$-exponential $e_*^{i k_1 H + i k_2 L_3}$. Let us define: $A= A_1 + A_2$, where:
\begin{equation}
A_1 = a I_{4 \times 4}, \hspace{0.5 cm} A_2 = c V = c \left(
\begin{array}{c c}
0 & \sigma_2\\
- \sigma_2 & 0
\end{array}
\right).
\end{equation}
Here, $a = \frac{i k_1}{2}$, $c= \frac{k_2}{2}$ and $\sigma_i$ $(i=1,2,3)$ are Pauli's spin matrices.

Now consider the matrices:
\begin{equation}
S_1 = \sqrt a I_{4 \times 4}, \hspace{0.5 cm} S_2 = \sqrt{\frac{c}{2}} \left(
\begin{array}{c c}
\sigma_2 & - I_{2 \times 2}\\
I_{2 \times 2} & \sigma_2
\end{array}
\right).
\end{equation}
They satisfy:
\begin{equation}
S_1^T S_1 = A_1 , \hspace{0.5 cm} S_2^T S_2 = A_2 , \hspace{0.5 cm} S_1^T S_2 + S_2^T S_1 =0.
\end{equation}
We conclude that $S_A = S_1 +S_2$. Moreover we have:
\begin{equation}
S_1 J S_2^T + S_2 J S_1^T =0.
\end{equation}
It then follows that $B= B_1 + B_2$, where:
\begin{equation}
B_1 = -a J, \hspace{0.5 cm} B_2 = c R = c \left(
\begin{array}{c c}
- \sigma_2 & 0\\
0 & - \sigma_2
\end{array}
\right).
\end{equation}
Consequently:
\begin{equation}
\left\{
\begin{array}{l l}
B_1^{2n} = (ia)^{2n} I_{ 4 \times 4 }, & B_1^{2n + 1} = i (ia)^{2n+1} J,\\
B_2^{2n} = c^{2n} I_{ 4 \times 4 }, & B_2^{2n + 1} = c^{2n+1} R,
\end{array}
\right.
\hspace{0.5 cm} n=0,1,2, \cdots
\end{equation}
This entails that:
\begin{equation}
\left\{
\begin{array}{l l}
\sin B_1 = - i \sin \left( \frac{k_1}{2} \right) J, &  \cos B_1 = \cos \left( \frac{k_1}{2} \right) I_{4 \times 4},\\
\sin B_2 = \sin \left( \frac{k_2}{2} \right) R, &  \cos B_2 = \cos \left( \frac{k_2}{2} \right) I_{4 \times 4}.
\end{array}
\right.
\end{equation}
It is easy to check that $B_1$ and $B_2$ commute. It then follows that:
\begin{equation}
\left\{
\begin{array}{l}
\sin B =  \sin B_1 \cos B_2 + \cos B_1 \sin B_2= - i \sin \left( \frac{k_1}{2} \right) \cos \left( \frac{k_2}{2} \right) J + \cos \left( \frac{k_1}{2} \right) \sin \left( \frac{k_2}{2} \right) R,\\
\cos B =  \cos B_1 \cos B_2 - \sin B_1 \sin B_2= \cos \left( \frac{k_1}{2} \right) \cos \left( \frac{k_2}{2} \right) I_{4 \times 4} + i \sin \left( \frac{k_1}{2} \right) \sin \left( \frac{k_2}{2} \right) V.
\end{array}
\right.
\end{equation}
To compute $\tan B$, we first need $\left( \cos B \right)^{-1}$:
\begin{equation}
\left( \cos B \right)^{-1} = \frac{\cos \left( \frac{k_1}{2} \right) \cos \left( \frac{k_2}{2} \right) I_{4 \times 4} - i \sin \left( \frac{k_1}{2} \right) \sin \left( \frac{k_2}{2} \right) V }{\cos^2 \left( \frac{k_1}{2} \right) \cos^2 \left( \frac{k_2}{2} \right) - \sin^2 \left( \frac{k_1}{2} \right) \sin^2 \left( \frac{k_2}{2} \right)}.
\end{equation}
We then get:
\begin{equation}
\tan B = \sin B \left( \cos B \right)^{-1} = \tan \left( \frac{k_1 + k_2}{2} \right) \left( \frac{R- i J}{2} \right) - \tan \left( \frac{k_1 - k_2}{2} \right) \left( \frac{R + i J}{2} \right).
\end{equation}
It remains to determine the inverse of $S_A$. First of all notice that:
\begin{equation}
A^{-1} = \frac{2}{k_2^2 - k_2^2} \left(
\begin{array}{c c}
i k_1 I_{2 \times 2} & - k_2 \sigma_2 \\
k_2 \sigma_2 & i k_1  I_{2 \times 2}
\end{array}
\right).
\end{equation}
Since $A= S_A^T S_A$, we conclude that $S_A^{-1} = A^{-1} S_A^T$. From this follows that $J S_A^{-1} J = - S_A^{-1}$.

Collecting all the results we get from (55):
\begin{equation}
\begin{array}{c}
\psi \left( k_1, k_2 \right) \equiv e_*^{i k_1 H + i k_2 L_3} = \frac{1}{\cos \left( \frac{k_1 + k_2}{2} \right) \cos \left( \frac{k_1 - k_2}{2} \right)} \times \\
\\
\times \exp \left\{i \tan \left( \frac{k_1 + k_2}{2} \right) \left( H + L_3 \right) + i \tan \left( \frac{k_1 - k_2}{2} \right) \left( H - L_3 \right) \right\}.
\end{array}
\end{equation}
Let us define $x = \frac{k_1 + k_2}{2}$, $y = \frac{k_1 - k_2}{2}$ and $\phi \left(x,y \right) = \psi \left(k_1 , k_2 \right)$. Again this is a distribution, periodic in the variables $x$, $y$, with periods $2 \pi$, and analytic for $x \ne (2n+1) \pi /2$, $y \ne (2m+1) \pi /2$ $(n,m \in {\cal Z} )$. Expanding in Fourier modes, we have:
\begin{equation}
\phi (x,y) = \sum_{n,m = - \infty}^{+ \infty} C_{n,m} e^{i n x+ i m y}.
\end{equation}
The Fourier modes are given by
\begin{equation}
C_{n,m} = \frac{1}{(2 \pi)^2} \int_{- \pi}^{\pi} dx \int_{- \pi}^{\pi} dy \hspace{0.2 cm} e^{- i n x - i m y} \phi (x,y) = \Pi_n \left(H + L_3 \right) \Pi_m \left(H - L_3 \right),
\end{equation}
where
\begin{equation}
\Pi_n (u) = \frac{1}{2 \pi} \int_{- \pi}^{\pi} dx \hspace{0.2 cm} \frac{e^{- i n  x + i u \tan x}}{\cos x} = \frac{1}{i \pi} \oint_C dz \hspace{0.2 cm} z^{-n} \left(z^2 +1 \right)^{-1} \exp \left[u \left(\frac{z^2 -1}{z^2 +1} \right) \right].
\end{equation}
We conclude that $\Pi_n (u)=0$ for $n<0$ or $n$ even, and that:
\begin{equation}
\Pi_{2n+1} (u) = 2 (-1)^n e^{-u} L_n (2u), \hspace{0.5 cm} n=0,1,2, \cdots
\end{equation}
We then have:
\begin{equation}
\begin{array}{c}
\Delta_* (H-E) * \Delta_* (L_3 - F) = \frac{1}{(2 \pi )^2} \int d k_1 \int d k_2 \hspace{0.2 cm} e_*^{ik_1 (H-E) + i k_2 ( L_3 - F)} =\\
\\
= \frac{2}{(2 \pi)^2} \int d x \int d y \hspace{0.2 cm} \phi (x,y) e^{-i(x+y) E - i (x-y) F}= \\
\\
= 2 \sum_{n,m=0}^{+ \infty} \Pi_{2n+1} (H+ L_3) \Pi_{2m+1} ( H-L_3) \delta (n-E -F) \delta (m - E + F)= \\
\\
=2 \sum_{r=0}^{+ \infty} \sum_{s \in I_r} \Pi_{r +s +1} (H+L_3) \Pi_{r -s +1} (H - L_3) \delta (E-E_r) \delta (F-F_s),
\end{array}
\end{equation}
where $r=n+m$, $s=n-m$, $E_r =r+1$, $F_s=s$ and $I_r = \left\{r,r-2,r-4, \cdots, 2-r, -r \right\}$. It is interesting to note that had we computed $\Delta_* (H-E)$, $\Delta_* (L_3 -F)$ separately, then the two spectra would have been independent. It is the $*$-product between the two expressions which forces the spectrum of $L_3$ to depend upon that of $H$.

Let us now compute the non-diagonal elements. The translation operator for $\hat L_3$ is
\begin{equation}
\hat T = \hat p_1^2- \hat p_2^2 + \hat q_1^2 - \hat q_2^2 + 2i \hat q_1 \hat q_2 + 2i \hat p_1 \hat p_2,
\end{equation}
which satisfies:
\begin{equation}
\left[\hat H, \hat T \right] =0, \hspace{0.5 cm} \left[\hat L_3, \hat T \right] = 2 \hat T.
\end{equation}
We then get from (18,34):
\begin{equation}
\Delta_* (H,L_3;E,E; F',F) = \frac{(-1)^n}{(2 \pi)^2} \int d \alpha \int d k_1 \int d k_2 \hspace{0.2 cm} \delta^{(n)} (\alpha) e_*^{i k_1 (H-E) + i k_2 (L_3 -F) + \mu \alpha T},
\end{equation}
where $F'=F +2n$ and $\mu (k_2) = \frac{2i k_2}{e^{2i k _2} -1}$. To compute the non-commutative exponential, let us define:
\begin{equation}
A= \left(
\begin{array}{c c}
a I_{2 \times 2} + i \gamma (\sigma_1 -i \sigma_3) & c \sigma_2\\
- c \sigma_2 &  a I_{2 \times 2} + i \gamma (\sigma_1 -i \sigma_3)
\end{array}
\right),
\end{equation}
where $a= \frac{ik_1}{2}$, $c= \frac{k_2}{2}$ and $\gamma = \alpha \mu (k_2)$. The matrix $S_A$ reads:
\begin{equation}
S_A = \left(
\begin{array}{c c}
-d & b\\
b & d
\end{array}
\right),
\end{equation}
where:
\begin{equation}
\left\{
\begin{array}{l}
b= b_0 I_{2 \times 2} + b_1 \sigma_1 - b_0 \sigma_2 + b_2 \sigma_3 ,\\
d= - b_0 I_{2 \times 2} - b_1 \sigma_1 + b_0 \sigma_2 + b_2 \sigma_3.
\end{array}
\right.
\end{equation}
Here, $b_1 = \frac{i \gamma}{4 b_0}$, $b_2 = \frac{c b_0}{\gamma}$ and $b_0$ is any solution of:
\begin{equation}
-16 c^2 b_0^4 + 8 \gamma^2  a b_0^2 + \gamma^4 =0.
\end{equation}
We then get from (56):
\begin{equation}
B = \left(
\begin{array}{c c}
- c \sigma_2 & a I_{2 \times 2} + i \gamma (\sigma_1 + i \sigma_3)\\
- a I_{2 \times 2} - i \gamma (\sigma_1 + i \sigma_3) & - c \sigma_2
\end{array}
\right),
\end{equation}
where we used $\sigma_i \sigma_j = \delta_{ij} I_{2 \times 2} + i \epsilon_{ijk} \sigma_k$. Let us write: $B= B_1 + B_2 + B_3$, with:
\begin{equation}
B_1 = - a J, \hspace{0.5 cm} B_2 = c R, \hspace{0.5 cm} B_3 = i \gamma \left(
\begin{array}{c c}
0 &  \sigma_1 + i \sigma_3\\
- \sigma_1 -  i \sigma_3 & 0
\end{array}
\right).
\end{equation}
We conclude that eqs.(133,134) still hold. Notice also that $B_2 B_3 + B_3 B_2=0$, $B_3^2 =0$. It then follows that:
\begin{equation}
e^{\pm i (B_2 + B_3)} = e^{\pm i B_2}  \pm i B_3 B_2^{-1} \sin B_2 ,
\end{equation}
where $B_2^{-1} =  \frac{2}{k_2} R$. Consequently:
\begin{equation}
\left\{
\begin{array}{l}
\sin (B_2 + B_3) =  \sin \left(\frac{k_2}{2} \right) \left(R +  \frac{2}{k_2}  B_3 \right) , \\
\\
cos (B_2 + B_3) = \cos B_2 = \cos \left(\frac{k_2}{2} \right) I_{4 \times 4}.
\end{array}
\right.
\end{equation}
Since $B_1$ and $B_2 + B_3$ commute, we get:
\begin{equation}
\begin{array}{c}
\sin B = \sin B_1 \cos (B_2 + B_3) + \cos B_1 \sin (B_2 + B_3) = \\
\\
= - i \sin \left( \frac{k_1}{2} \right) \cos \left( \frac{k_2}{2} \right) J + \cos \left( \frac{k_1}{2} \right)  \sin \left( \frac{k_2}{2} \right) \left( R + \frac{2}{k_2} B_3 \right),
\end{array}
\end{equation}
and
\begin{equation}
\begin{array}{c}
\cos B = \cos B_1 \cos (B_2 + B_3) - \sin B_1 \sin (B_2 + B_3) = \\
\\
= \cos \left( \frac{k_1}{2} \right) \cos \left( \frac{k_2}{2} \right) I_{4 \times 4} + i \sin \left( \frac{k_1}{2} \right) \sin \left( \frac{k_2}{2} \right) \left( V + \frac{i \gamma}{c} Z \right),
\end{array}
\end{equation}
where:
$$
Z= \left(
\begin{array}{c c}
\sigma_1 + i \sigma_3 & 0\\
0 & \sigma_1 + i \sigma_3
\end{array}
\right),
$$
and $V$ was defined in eq.(128). A lengthy calculation then yields:
\begin{equation}
\begin{array}{c}
\left( \cos B \right)^{-1} = \frac{1}{\cos \left(\frac{k_1 + k_2}{2} \right) \cos \left(\frac{k_1 - k_2}{2} \right)} \times \left\{\cos \left(\frac{k_1}{2} \right) \cos \left(\frac{k_2}{2} \right) I_{4 \times 4}  \right.\\
\\
\left. - i \sin \left(\frac{k_1}{2} \right) \sin \left(\frac{k_2}{2} \right) \left( V + \frac{i \gamma}{c} Z \right) \right\},
\end{array}
\end{equation}
and:
\begin{equation}
\tan B =  \tan \left( \frac{k_1 + k_2}{2} \right)  \left(\frac{R - iJ}{2} - \frac{1}{k_2} B_3 \right) - \tan \left( \frac{k_1 - k_2}{2} \right)  \left(\frac{R + iJ}{2} - \frac{1}{k_2} B_3 \right).
\end{equation}
As before: $\left( \det \cos B \right)^{1/2} = \cos \left( \frac{k_1 + k_2}{2} \right) \cos \left( \frac{k_1 - k_2}{2} \right)$.
Finally:
\begin{equation}
A^{-1} = \frac{1}{a^2 + c^2} \left(
\begin{array}{c c}
a I_{2 \times 2} - i \gamma (\sigma_1 - i \sigma_3) & - c \sigma_2\\
c \sigma_2 & a I_{2 \times 2} - i \gamma (\sigma_1 - i \sigma_3)
\end{array}
\right), \hspace{0.5 cm} S_A^{-1} = A^{-1} S_A^T,
\end{equation}
and $J S_A^{-1} J = S_A^{-1}$. Substituting these results in eqs.(55,147), we get:
\begin{equation}
\Delta_* (H, L_3; E, E ; F', F) = \frac{1}{2 \pi^2} \int dx \int dy \hspace{0.2 cm} e^{-i (x+y) E - i (x-y) F} \psi^{(n)} (x,y),
\end{equation}
where
\begin{equation}
\left. \psi^{(n)} (x,y) = \frac{1}{\cos x \cos y} \frac{\partial^n}{\partial \alpha^n} \left\{i (H+ L_3) \tan x + i (H-L_3) \tan y + \alpha  T (\tan x + i) (\tan y - i) \right\} \right|_{\alpha =0},
\end{equation}
is a periodic distribution in the variables $x$, $y$ with periods $2 \pi$, and analytic for $x \ne (2l+1) \pi /2$, $y \ne (2m+1) \pi /2$ $(l , m \in {\cal Z})$. Again we expand it in a Fourier series:
\begin{equation}
\psi^{(n)} (x,y) = \sum_{l,m= - \infty}^{+ \infty} C_{l,m}^{(n)} e^{ilx + imy},
\end{equation}
where
\begin{equation}
C_{l,m}^{(n)} = \frac{1}{(2 \pi)^2} \int_{- \pi}^{\pi} dx \int_{- \pi}^{\pi} dy \hspace{0.2 cm} \psi^{(n)} (x,y) e^{- i l x - i my} = \frac{T^n}{(2 \pi)^2} I_l^{(n)} (H+ L_3) J_m^{(n)} (H-L_3).
\end{equation}
Here,
\begin{equation}
\begin{array}{c}
I_l^{(n)} = \int_{- \pi}^{\pi} dx \frac{(\tan x +i)^n}{\cos x} \exp \left[i(H+L_3) \tan x - i l x \right] = \\
\\
=- (2i)^{n+1} \oint_C  \hspace{0.2 cm} dz z^{-l} (z^2 +1)^{-n-1} \exp \left[(H+L_3) \left(\frac{z^2-1}{z^2 +1} \right) \right].
\end{array}
\end{equation}
Again, this vanishes if $l <0$ or if $l$ is even. We are left with:
\begin{equation}
I_{2l+1}^{(n)} =(2i)^{n+1} (-1)^{l +1} 2\pi i e^{- (H+L_3)} L_l^n (2H +2 L_3), \hspace{0.5 cm} l=0,1,2,\cdots
\end{equation}
Likewise, the nonvanishing terms of $J_m^{(n)}$ are given by:
\begin{equation}
\begin{array}{c}
J_{2m+1}^{(n)} = \int_{- \pi}^{\pi} dy \hspace{0.2 cm} \frac{(\tan y -i)^n}{\cos y} \exp \left[i (H-L_3) \tan y - i (2m+1)y \right] =\\
\\
= (-2i)^{n+1} e^{-(H-L_3)} 2 \pi i (-1)^{m-n} L_{m-n}^n (2 H -2 L_3), \hspace{0.5 cm} m \ge n.
\end{array}
\end{equation}
We then get:
\begin{equation}
\begin{array}{c}
\Delta_* (H,L_3; E, E; F',F) = \frac{1}{2 \pi^2} \sum_{l=0}^{+ \infty} \sum_{m=n}^{+ \infty} \int dx \int dy \hspace{0.2 cm} C_{2l+1, 2m+1}^{(n)} e^{-i x (E+F-2l-1) - i y (E-F-2m-1)}=\\
\\
= \sum_{r=n}^{+ \infty} \sum_{s \in J_r} \delta (E-E_r) \delta (F-F_s) C_{r+s+1,r-s+ 1}^{(n)},
\end{array}
\end{equation}
where $J_r = \left\{-r, -r+2, -r+4, \cdots, t_r \right\}$ and $t_r= r-n$ (for $n$ even) or $r-n-1$ (for $n$ odd).

Using the fact that $\left[ \hat L_3, \hat T \right] = 2 \hat T$ and $\left[\hat T , \hat T^{\dagger} \right] = 16 \hat L_3$, we can prove that:
\begin{equation}
|-r + 2l> = \sqrt{\frac{(r-l)!}{(r- l')!}} \left(\frac{\hat T}{4} \right)^{l-l'}  |-r + 2l'>, \hspace{0.5 cm} l \ge l'.
\end{equation}
This means that (cf.(29)):
$$
\beta_{-r+l+l',-r+2l} =  \sqrt{\frac{(r-l)!}{(r- l')!}} \times \left( \frac{1}{4} \right)^{l - l'}.
$$
The non-diagonal Wigner functions $F_{rs',rs}^W$ can be read off from eqs.(168,169)) for $s'> s$. To obtain $F_{rs',rs}^W$ for $s'<s$, we complex conjugate the latter expression and perform the substitution $s' \longleftrightarrow s$. The general result (for arbitrary $m$ and $\omega$ and restoring the $\hbar$'s) can then be written in compact form:
\begin{equation}
\begin{array}{c}
F_{rs',rs}^W = \frac{(-1)^r}{(\pi \hbar)^2 \omega} \left[\frac{\left(\frac{r-s}{2} \right)!}{\left(\frac{r- s'}{2} \right)!} \right]^{sign (s-s')/2} \left(- \frac{T}{\hbar m \omega} \right)^{\frac{|s-s'|+s-s'}{4}} \left( - \frac{T^{\dagger}}{\hbar m \omega} \right)^{\frac{|s-s'|+s'-s}{4}} e^{- \frac{2H}{\omega \hbar}} \times \\
\\
\times L_{\frac{2r+s+s'-|s-s'|}{4}}^{\frac{|s-s'|}{2}} \left( \frac{2(H+ \omega L_3)}{\hbar \omega} \right) L_{\frac{2r- s - s'-|s-s'|}{4}}^{\frac{|s-s'|}{2}} \left( \frac{2(H - \omega L_3)}{\hbar \omega} \right),
\end{array}
\end{equation}
with $s,s' \in I_r = \left\{-r, -r+2, \cdots , r-2, r \right\}$. A lengthy calculation shows that:
\begin{equation}
\left\{
\begin{array}{l}
H*F_{rs',rs}^W=F_{rs',rs}^W*H = E_r F_{rs',rs}^W,\\
\\
L_3* F_{rs',rs}^W = F_{s'} F_{rs',rs}^W, \hspace{0.5 cm} F_{rs',rs}^W *L_3 = F_s F_{rs',rs}^W,
\end{array}
\right.
\end{equation}
with $E_r = \omega \hbar \left( r + \frac{1}{2} \right)$ and $F_s = s \hbar$.

\section{Conclusions}

Let us briefly summarize our results. We derived a formal solution for an arbitrary (diagonal and non-diagonal) $*$-genvalue equation. The formalism was designed to incorporate observables with discrete and continuous spectra as well as multidimensional systems. This formal solution was interpreted as a non-commutative generalization of the Dirac delta distribution, identifying $*$-hypersurfaces in the non-commutative phase space. Moreover, we presented the complete specification of Wigner quantum mechanics in the Heisenberg picture. We further derived the explicit expression of our formal solution for an arbitrary quadratic phase space functional. If the Hessian matrix of this functional is proportional to a symplectic matrix, then the aforementioned explicit expression simplifies drastically and we can systematically analyze the continuous and discrete spectrum cases. Finally, we apply our formalism to the one- and two-dimensional harmonic oscillators and to the linear potential. The diagonal terms were previously solved in the literature (\cite{Fairlie1}, \cite{Dahl1}, \cite{Flato1}) using other methods. The non-diagonal terms are to the best of our knowledge new.

\paragraph*{Acknowledgments.}

We would like to thank Aleksandar Mikovic and Cosmas Zachos for useful suggestions.
This work was partially supported by the grants POCTI/MAT/45306/2002 and POCTI/FNU/49543/2002.

\end{document}